\pdfoutput=1
\documentclass[colorinlistoftodos]{article}

\usepackage{amsfonts}
\usepackage{amsmath}
\usepackage{amsthm}
\usepackage{bm}
\usepackage{algorithm}
\usepackage{algorithmic}

\usepackage{pdfpages}

\usepackage{nicefrac}
\usepackage{float}

\usepackage{pgfplots}
\pgfplotsset{compat=1.14}
\usepgfplotslibrary{patchplots}

\usepackage{environ}
\usepackage[utf8x]{inputenc} 

\usepackage{mdframed}
\usepackage{dsfont}

\usepackage{bibentry}
\usepackage{subfigure}
\usepackage{placeins}

\usepackage[hidelinks]{hyperref}

\usepackage{mdframed,framed,color,verbatim}
\definecolor{shadecolor}{rgb}{.9, .9, .9}
\definecolor{outlinecolor}{rgb}{0.7,0.7,0.9}

\newtheorem{lem}{Lemma}
\newtheorem{thm}{Theorem}

\theoremstyle{definition}
\newtheorem{de}{Definition}
\theoremstyle{remark}
\newtheorem{remark}{Remark}
\theoremstyle{definition}
\newtheorem{ts}{Test set}
\theoremstyle{definition}
\newtheorem{question}{Question}

\newcommand{\rank}{Rank}
\newcommand{\rankof}[1]{\rank \left( #1 \right)}
\newcommand{\paper}{{work}}

\newcommand{\nhalf}{\nicefrac{1}{2}}
\newcommand{\lognk}{\log \frac{n}{k}}
\newcommand{\D}{\mathcal{D}}
\newcommand{\notD}{\left[n\right]\setminus\D}

\newcommand{\prmid}{\;\middle\vert\;}

\newcommand{\algS}{{S^\sharp}}
\newcommand{\algSk}{{S^\sharp_k}}
\newcommand{\algSt}{{S^\sharp_t}}
\newcommand{\algSz}{{S^\sharp_0}}
\newcommand{\algStmo}{{S^\sharp_{t-1}}}
\newcommand{\algSH}{{S_{H,S}}}
\newcommand{\algSTH}{{S_{H,T}}}

\newcommand{\SA}{I(A)}

\newcommand{\ASpace}{\{0,1\}^{m \times n}}

\newcommand{\tub}{\left(1+o(1)\right)2k\frac{\lognk}{\log k}} 
\newcommand{\Mqgt}{M_{QGT}}
\newcommand{\hMqgt}{\hat{M}_{QGT}}

\makeatletter
\newcommand{\whp}{w.h.p\@ifnextchar.{}{.\@{ }}}
\newcommand{\wlogA}{w.l.o.g\@ifnextchar.{}{.\@{ }}}
\newcommand{\etc}{etc\@ifnextchar.{}{.\@{ }}}
\makeatother

\newcommand{\one}{\mathds{1}}

\usepackage{subfiles} 
\newcommand{\main}{.}

\title{Quantitative Group Testing and the rank of random matrices}
\author{Uriel Feige \thanks{The Weizmann Institute of Science, Israel. Email: {\tt uriel.feige@weizmann.ac.il}.}\and Amir Lellouche \thanks{The Weizmann Institute of Science, Israel. Email: {\tt amir.lellouche@weizmann.ac.il}.}
}
\begin{document}
	\maketitle
	
	\begin{abstract}
		Given a random Bernoulli matrix $ A\in \{0,1\}^{m\times n} $, an integer $ 0< k < n $ and the vector $ y:=Ax $, where $ x \in \{0,1\}^n $ is of Hamming weight $ k $, the objective in the {\em Quantitative Group Testing} (QGT) problem is to recover $ x $.  This problem is more difficult the smaller $m$ is.
		For parameter ranges of interest to us, known polynomial time algorithms require values of $m$ that are much larger than $k$.
		
		In this work, we define a seemingly easier problem that we refer to as {\em Subset Select}.
		Given the same input as in QGT, the objective in Subset Select is to return a subset $ S \subseteq [n] $ of cardinality $ m $, such that for all $ i\in [n] $, if $ x_i = 1 $ then $ i\in S $.
		We show that if the square submatrix of $A$ defined by the columns indexed by $S$ has nearly full rank, then from the solution of the Subset Select problem we can recover in polynomial-time the solution $x$ to the QGT problem.
		We conjecture that for every polynomial time Subset Select algorithm, the resulting output matrix will satisfy the desired rank condition.
		We prove the conjecture for some classes of algorithms.
		Using this reduction, we provide some examples of how to improve known QGT algorithms.
		Using theoretical analysis and simulations, we demonstrate that the modified algorithms solve the QGT problem for values of $ m $ that are smaller than those required for the original algorithms.
	\end{abstract}

	\section{Introduction}\label{sec:introduction}
	 
		Quantitative Group Testing (QGT) is the problem of detecting $k$ defective items among a total of $ n $ items by performing tests on $ m $ different pools.
		We refer to elements of $ [n] $ as items, and those which are in the subset $ \D \subseteq [n] $ as defective items. 
		A test consists of a pool of items, and its \emph{outcome} is the number of defective items belonging to that pool.
		Formally, we define the pools as a \emph{test matrix} $ A \in \{0,1\}^{m \times n} $ such that item $ i\in[n] $ is in pool $ j\in[m] $ if and only if  $ A_{ji} = 1 $.
		Denote by $ x \in \{0,1\}^n $ the indicator vector of the subset $ \D $, that is $ x_i=1 $ if and only if $ i \in \D $.
		We denote the \emph{tests outcome} with $ y \in \mathbb{Z}_+^m $ so that $ y_j = t $ if in the $ j $th pool there were $ t $ defective items. Namely,
		\begin{equation*}
			y:=Ax.
		\end{equation*}
		
		We consider the setting in which the test pools construction is random, and the algorithm has no control over them. 
		Specifically, the probability of every item to be in a test pool is $ \nhalf $, independently. 
		It is then that $ A \in \ASpace $ is a random Bernoulli matrix (that is, each entry is i.i.d.\@ Bernoulli random variable with $ p=\nhalf $).
		To conclude, given $ A, k$ and $ y $  the QGT problem is to recover $ \D $.
		
		QGT comes from a family of problems that has practical applications. 
		For more details, see Section \ref{sec:related-works}.
		This problem is also of theoretical interest;
		determining the minimum number of tests such that \whp the defective items can be efficiently recovered (namely, by a polynomial time algorithm) is a challenging open problem.
		
		In our work we encounter  other interesting questions that are new in the QGT problem context, and may be of independent interest.
		They relate to the rank distribution of a random Bernoulli matrix or its submatrices.
		For more details see Section \ref{sec:discussion}.
					
		It is known that a random square Bernoulli matrix is nonsingular with overwhelming probability; 
		therefore, if $ m\ge n $, one can find the defective items $ \D $ by solving the system of linear equations $ Ax=y $. 
		Interestingly, it is possible to solve \emph{with high probability} (w.h.p.) the QGT problem also when the number of tests is much smaller than the number of items. Henceforth we will only consider $ m <n $.
		
		When $ k $ is assumed to be linear in $ n $, we refer to such setting as the \emph{linear regime}. 
		In the \emph{sublinear regime}, $ k = n^\theta $ is assumed for some constant $ \theta \in (0,1) $.

		Denote by $ \Mqgt=\Mqgt(n,k)\in \mathbb{R}_+ $ the \emph{information-theoretic} threshold of the QGT problem. 
		Formally, $ \Mqgt(n,k) $ is the minimal integer such that for all $ m > \Mqgt $ the system of linear equation $ Az=y $ has \whp a single solution $ z\in \{0,1\}^n$ of Hamming weight $ k $. 
		It is known that
		\begin{equation}\label{intro:eq:info-theortic-bound-linear}
			(1 - o(1))\frac{2n}{\log k} h\left (\frac{k}{n}\right ) 
			\le \Mqgt 
			\le (1 + o(1))\frac{2n}{\log k} h\left (\frac{k}{n}\right ),
		\end{equation}
		where $ h $ is the binary entropy function defined as $ h(p) = -p\log (p) - (1-p)\log (1-p)$.		
		The two inequalities of (\ref{intro:eq:info-theortic-bound-linear}) hold in the linear regime as an immediate corollary of \cite{scarlett2017phase} and \cite{alaoui2019decoding}, respectively.
		In the {sublinear regime} the above bounds read as 
		\begin{equation}\label{intro:eq:info-theortic-bound-sublinear}
			(1 - o(1))2k\frac{\lognk}{\log k} 
			\le \Mqgt 
			\le (1 + o(1))2k\frac{\lognk}{\log k}.
		\end{equation}		
		The first inequality is a consequence of \cite{lindstrom1975determining,djackov1975search}. 
		As these papers are hard to find, we provide a proof sketch in the Appendix (see Section \ref{appendix:tlb}) for this inequality.
		The second inequality is a known result and for completeness we provide a proof for it in the Appendix (see Section \ref{appendix:tub}, and also \cite{gebhard2019quantitative} for a similar proof). Note that in the sublinear regime $ \frac{\lognk}{\log k}=\frac{1-\theta}{\theta} $ is  constant.
		
		However, there is no known polynomial-time algorithm that meets these information-theoretic bounds.
		As we survey in section \ref{sec:related-works}, all known  algorithms (to the best of our knowledge) that solve the QGT problem efficiently require at least a factor of $ \Omega(\log k) $ more tests. 
		Generally, studies of the algorithmic aspect focus on the leading constants of the required number of tests and sometimes also on the time complexity of the algorithm. In this work we focus on both of these matters.

	\subsection{Notation}	
			
			Consider a matrix $A \in \{0,1\}^{m \times n}$ and a subset $S \subseteq [n]$.
			We use the notation $A|_S$ to refer the submatrix of $A$ induced by columns with an index in $ S $, so that $A|_S \in \{0,1\}^{m \times |S|}$. 
			We denote the complementary subset of $ S $ with $ \bar{S}$, formally $ \bar{S}:=[n]\setminus S $.
			We use the notation $ \one_S \in \{0,1\}^n $ to refer the indicator vector of $ S $, that is $ (\one_{S})_i = 1 $ if $ i \in S $ or $ (\one_{S})_i = 0 $ otherwise; for instance $ x=\one_{\D} $.
			We use the notation $ 1_N $ to denote a vector in $ \mathbb{R}^N $ that all of its entries are one, namely $ 1_N:=\one_{[N]} $. For some integers $ M,N $ we denote by $ 1_{M\times N} \in \mathbb{R}^{M \times N}$ the matrix that all of its entries are one.
			For a set of vectors $ S = \{v_1, v_2, \ldots, v_n\}\subseteq \mathbb{R}^m $, denote by $ \langle S \rangle $ the linear space spanned by them.
			We denote the base $ 2 $  logarithm with $ \log $ and the base $ e $ logarithm with $ \ln $.
	
	\subsection{Our Contribution}\label{sec:contributions}
	
	\subsubsection{The Subset Select problem}
		We provide a new family of algorithms to solve the QGT problem. 
		They are based on a new problem that we introduce and call \emph{Subset Select}.
		Our aim is to reduce the number of required tests for known algorithms that solve the QGT problem without adding too much running time. 
		
		We define the Subset Select problem as an attempt to relax the QGT problem.
		Given a QGT problem instance $ A,k $ and $ y $, the objective of Subset Select is to output a subset $ \algS \subseteq [n] $ of cardinality at most $ m $ such that $ \D \subseteq \algS $. 
		Recall that QGT problem outcome needs to satisfy $ \algS = \D $. 
		Therefore, we can think of the Subset Select problem as a relaxed variation that allows some false positives in the output.	
		However, this is not always the case. 
		
		Note that the Subset Select problem is infeasible when $ k > m $. 
		Therefore, it is a relaxation of the QGT problem only when the number of defective items is smaller than the amount of tests.
		Unfortunately, unless we consider the sublinear-regime with $ \theta < \frac{2}{3} $, we have that $ k>\Mqgt $.
		This implies that the method we present in this {\paper} will not lead to a breakthrough with finding an algorithmic threshold that meets the information-theoretic one.
		However, it may still improve the known algorithmic thresholds.
		As discussed before, all known efficient algorithms require a factor of $ \Omega(\log k) $ more tests than the information theoretic threshold.
		Now, observe that if $ k>\frac{n}{2} $, we may switch labels between defectives and non-defectives and considering a different tests outcome $ y':= k\cdot1_m - y $. 
		Therefore, we may assume  w.l.o.g.\@ that $ k \le \frac{n}{2} $.	
		With this assumption and as we survey in Section \ref{sec:related-works}, for all known algorithms $ k < m $, and in the sublinear regime we even have that $ k=o(m) $. 
			
		Consider an efficient algorithm $ ALG $ for the Subset Select problem and denote its outcome as $ \algS $, with $ |\algS| = m $.
		If the submatrix $ A|_{\algS} $ has $ \rankof{A|_{\algS}}=|\algS| $ and additionally $ \D \subseteq \algS $ then by solving the system of linear equations $ (A|_{\algS})z=y $ one can recover $ \D $.
		
		If the entries of $ A $ were, for example, i.i.d.\@ Gaussian random variables then we would have with probability $ 1 $ that $ \rankof{A|_{\algS}} = |\algS|$ for any such subset $ \algS $.
		However, in our setting this is not the case. 
		As an easy example, there exists, with overwhelming probability, a row $ j\in[m] $ in $ A $ such that at least $ \frac{n}{2} $ of its entries are zero.  
		Therefore, if $m \le \frac{n}{2}$ (which holds in cases that interest us), there exists a subset $ S \subseteq [n] $ of cardinality $ m $ such that for all $ i \in S $ we have $ A_{ji} = 0 $.
		Hence $ \rankof{A|_S}<|S| $. 
		
		In order to use algorithms that solve Subset Select in the context of QGT problem we first relax the full rank requirement. 
		That is, in Theorem \ref{thm:SSA-rank-condition} we show that if $A|_S$ is of rank $ |S| -O(\log n) $ then we can also recover the defective items in polynomial-time. 
		
		\begin{thm}		\label{thm:SSA-rank-condition}
			Consider an algorithm $ ALG $ for the Subset Select problem that runs in time $ T_{ALG} $ and denote its outcome as $ \algS $.
			Suppose that there exists a constant $ C\ge0 $ and some $ M:=M(n,k)\in \mathbb{R} $ such that for all $ m > M $, the following properties hold w.h.p.:
			\begin{enumerate}
				\item The \emph{containment} property: the algorithm succeeds, i.e., $ \D \subseteq \algS $.
				\item The \emph{relaxed rank} property: the submatrix $ A|_\algS $ is of rank at least $ |\algS|-C\log n $.
			\end{enumerate}
			Then there exists an algorithm that solves the QGT problem \whp when $ m>\max(M,M_{QGT}) $.
			
			Furthermore, this algorithm runs in time $O(m^3 +n^C m \log n) + T_{ALG}$.
		\end{thm}
	
		Consider a random Bernoulli matrix $ A\in\{0,1\}^{m \times n} $ where $ m $ is a function of $ n $. 
		We conjecture that for a sufficiently large $ n $, all of $ A $'s submatrices 
		induced by $ m $ columns (i.e., submatrices of the form $  A|_S $, for some $ S\subseteq[n] $ of cardinality $ m $) have \whp rank of at least $ m-O(\log n) $.
		If the conjecture holds, then every Subset Select algorithm has the relaxed rank property.
		Unfortunately, we do not know if the conjecture holds.
		The main focus in this {\paper} is then to prove for some special cases that a Subset Select algorithm has the relaxed rank property.
	
	\subsubsection{Algorithms for the Subset Select problem}
		\label{sec:contributions:ssa-algorithms}
		The first QGT algorithm that we study is the \emph{$k$-Thresholding}	
		algorithm (see Algorithm \ref{alg:Basic-Thresholding}).
		This is a variation of the \emph{Basic-Thresholding} algorithm \cite{foucart2013invitation} that uses some properties of the QGT problem setting.
		For every item $ i \in [n] $, its score $ \psi_i $ is a function (that we define later in section \ref{boosting:subsec:basic-thresholding}) of the $ i $th column of the test-matrix and the test-outcome-vector $ y $;
		high score value for an item $ i $, indicates that item $ i $ is likely to be defective.
		The score $ \psi_i $ is slightly different than the one that is usually considered. 
		In \cite{foucart2013invitation}, the score $ \phi_i $ of item $ i $ is the normalized sum over the outcomes of all tests in which item $ i $ participates. Namely, $ \phi_i:=\sum_{j=1}^m A_{ji}y_j / |A_i|$.
		In the score function of the $ k $-Thresholding algorithm $ \psi_i $ we use the fact that $ k $ is known (or can be computed, see Section \ref{section:k-is-unkown-or-random}) to include information also from test pools which item $ i $ does not participate in. 
		Therefore this new score function $ \psi_i $ uses information from almost twice as many tests. Moreover, using this score function also facilitates the analysis.
		\begin{algorithm}
			\caption{$k$-Thresholding}\label{alg:Basic-Thresholding}
			\begin{algorithmic}
				\STATE \textbf{Input:} $ A,y,k $
				\begin{enumerate}
					\item For every item $i \in [n]$, calculate its score $\psi_i$.
					\item Sort the items in descending order of score. 
					Return the subset $\algS \subseteq [n]$ that contains the first $k$ items.
				\end{enumerate}
			\end{algorithmic}
		\end{algorithm}	
		
		We adapt this algorithm in a straightforward way to the Subset Select problem.
		Instead of returning the first $ k $ items sorted by score, the modified algorithm returns the first $ m $ items. 
		We call this variation the \emph{$m$-Thresholding} algorithm.			
		We prove in Lemma \ref{boosting:lem:basic-thresholding-is-consistent} that  this algorithm has the relaxed rank property. Therefore, by Theorem \ref{thm:SSA-rank-condition}, $m$-Thresholding yields an algorithm for the QGT problem.
		
		In Lemma \ref{boosting:lem:top-k-number-of-tests} and Lemma \ref{boosting:lem:top-m-number-of-tests}, we give thresholds on the required number of tests such that $k$-Thresholding and $m$-Thresholding \whp succeed, respectively.
    Between the two upper-bounds, there is a gap in the latter algorithm's favor.
		This gap is of a constant factor that depends on the ratio between $ k $ and $ n $ (see Figure \ref{figure:bt-ssbt-theory}).
		Our empirical simulations results agree with theoretical calculations (see test set \ref{simulations:testset:bt-ssbt-2kssbt} in section \ref{sec:empirical}).
		
		With a motivation to improve the required number of tests for generic QGT algorithms using a similar method, we provide the following framework. 
		Given an algorithm $ ALG $ for the QGT problem, this framework constructs a new algorithm \emph{$ (ALG) $-Then-Thresholding} (see Algorithm \ref{alg:ALG-Then-Thresholding}) for the Subset Select problem. 
		For this framework to be useful we need to address two matters in the resulting Subset Select algorithm. 
		First, we define the \emph{threshold} $ M(n,k)\in \mathbb{N} $ of a QGT (resp.\@ Subset Select) algorithm $ ALG $ to be the minimal integer such that for all $ m>M(n,k) $, $ ALG $ returns \whp a subset $  \algS $ with $ \algS = \D $ (resp.\@ $ \D \subseteq \algS$).
		To construct an algorithm that requires fewer tests,
		we want the threshold of the QGT algorithm $ ALG $ to be larger than the threshold of the resulting Subset Select algorithm.
		Second, it is important to ensure that the resulting Subset Select algorithm can indeed can be used to solve efficiently the QGT problem (e.g, if the resulting Subset Select algorithm has the relaxed rank property, we can use Theorem \ref{thm:SSA-rank-condition}).
		Both of these matters strongly depend on the QGT algorithm $ ALG $.
		However, in Lemma \ref{boosting:lem:when-is-second-stage-consistent} we provide a sufficient condition on $ ALG $ so that the resulting Subset Select algorithm will have the relaxed rank property.
		
		To examine which algorithms are improved by this framework  in terms of the required amount of tests, we conducted simulations and documented for each algorithm the ratio between the number of tests and its success frequency for various values of $ n $ and $ k $ (see test set \ref{simulations:testset:tbt-framework} in section \ref{sec:empirical}).
		In some cases, the simulation results showed a significant improvement of the resulting algorithm versus its original.
		However, among the algorithms simulated, the one requiring the smallest number of tests (though also the one most demanding in terms of computational resources) was Bin-BP (see Section \ref{sec:related-works}), and for this algorithm the improvements offered by our framework appear to be insignificant.

		\begin{algorithm}
			\caption{($ ALG $)-Then-Thresholding} \label{alg:ALG-Then-Thresholding}
			\begin{algorithmic}
				\STATE \textbf{Input:} $ A,y,k $
				\begin{enumerate}
					\item Evaluate $ ALG $ on the problem instance $ S \leftarrow ALG(A,y,k) $.
					\item Get the residual $ r \leftarrow y - A\one_S $.
					\item For every item $i \in [n] \setminus S $, calculate its score $\phi_i := \langle A_i, r \rangle / \|A_i\|_1$.
					\item  Sort the items in descending order of score. Define the subset $S' \subseteq [n]\setminus S $ that contains the first $m-k$ items.
					\item Return the subset $ \algS = S'\cup S $.
				\end{enumerate}
			\end{algorithmic}
		\end{algorithm}	
	
		We believe that every polynomial time algorithm that solves the Subset Select problem will have the relaxed rank property and hence will lead to an algorithm for the QGT problem.		
		As we do not know if this is true, we provide also the following theorem.
		\begin{thm} \label{thm:SSA-with-little-order-term}
			If $ALG$ is an algorithm that solves the Subset Select problem efficiently and \whp for $ m>M\ge\Mqgt $ where $ M=M(n,k) \in \mathbb{R} $, then
			there exists an algorithm $ ALG' $ that solves the QGT problem efficiently and \whp for $ m>(1+o(1))M $.
			
			Furthermore, if $ ALG $ runs in time $ T_{ALG} $ then $ ALG' $ runs in time $ T_{ALG} + O(m^3) $.
		\end{thm}
		\begin{remark}
			The $ m^3 $ term can be replaced by $ m^\omega $, where $ \omega \le 3 $ is the matrix multiplication constant.
		\end{remark}

	\subsection{Related work}\label{sec:related-works}
	 
		The QGT problem is a special case of a more general problem, referred to as \emph{Compressed Sensing}.
		In Compressed Sensing the objective is to recover a \emph{signal} $ x\in \mathbb{R}^n $ given $ m $ linear measurements $ y\in \mathbb{R}^m $ such that $ y:=Ax $ for some given \emph{sensing matrix} $ A \in \mathbb{R}^{m\times n} $.
		If $ A $ has  full rank, then the signal $ x $ can be recovered.
		However, if the signal $ x $  is \emph{$ k $-sparse} for some $ k < n $ (a vector $ x $ is $ k $-sparse if it has $ k $ non-zero entries ,i.e., $ \|x\|_{0}=k $), then lower rank may suffice.
		Note that in contrast with the QGT problem, in Compressed Sensing we do not require that $ x \in \{0,1\}^n $ and $A\in \{0,1\}^{m\times n} $.
		
		The sensing matrix $ A \in \mathbb{R}^{m\times n} $ is assumed to be random in most results in this field.
		The most prominent distributions of the sensing matrix handled in the literature are the \emph{Gaussian} and the \emph{Rademacher} distributions.
		We say that a matrix $A \in \mathbb{R}^{m \times n}$ is Gaussian if its entries are independent and identically distributed standard normal random variables (to be more precise, usually the normalization of this matrix is considered, that is $A'=\frac{1}{\sqrt{m}} A$). 
		A Rademacher matrix $A \in \{\pm 1\}^{m \times n}$ has its entries independently $-1$ or $1$ with equal probability $\nhalf$ (again, usually the normalized matrix is considered).
		
		Results regarding Rademacher sensing matrix can be used in our setting.
		Recall, that in the QGT problem we consider a Bernoulli random test matrix.
		Nonetheless, $k$ is known (or can be computed in the region of interest, for details see section \ref{section:k-is-unkown-or-random}). 
		Therefore, it is possible to reduce the QGT problem instance $ (A,y,k) $ to compressed sensing with the Rademacher sensing matrix by computing $(1_{m\times n}-2A) \in \{\pm 1\}^{m \times n}$ and $k\cdot 1_{m} - 2y$ as the Rademacher matrix and its measurements of the signal.
		
		In \cite{Rauhut2010Nonuniform} (see also \cite{foucart2013invitation}), they show that the recovery of an arbitrary k-sparse signal with a Rademacher matrix can be made \whp for 
		\begin{equation*}
			m\geq (1+o(1))2k\ln n
		\end{equation*}
		tests and when $ k \gg 1 $ via \emph{Basis Pursuit} (BP) algorithm. That is, return the following optimization problem solution:
		\begin{equation*}
					\min\|z\|_1 \ \text{subject to}\ Az=y\ \text{for}\ {z\in \mathbb{R}^n}
		\end{equation*}
		Some other studies, such as \cite{stojnic2009simple, donoho2009counting, kueng2017robust}, consider a positive prior of the signal (i.e., x is a k-sparse non-negative vector), and therefore study the modified optimization problem, that we refer to as the \emph{P-BP}:
		\begin{equation*}
					\min\|z\|_1 \ \text{subject to}\ Az=y\ \text{for}\ {z\in \mathbb{R}^n_+}
		\end{equation*}
		The QGT problem has also been addressed directly in compressed sensing with Rademacher or Bernoulli sensing matrix together with a binary prior of the signal. In \cite{keiper2017compressed,stojnic2009simple} they analyze this optimization problem:
		\begin{equation} \label{alg:BP-bin}
			\min\|z\|_1 \ \text{subject to}\ Az=y\ \text{for}\ {z\in [0,1]^n}.
		\end{equation}
		We refer to (\ref{alg:BP-bin}) as the \emph{BIN-BP} algorithm.
		In \cite{flinth2018recovery} they show that asymptotically for $ m>O(k \lognk) $ (with a large constant factor), there is a single solution $ z $ that satisfies the above constrains and therefore, they consider a least square variation
		\begin{equation*}
			\min\|y-Az\|_2 \ \text{for}\ {z\in [0,1]^n}.
		\end{equation*}
		The above formulation is aimed to handle noise in the measurements vector $ y $.  	
		We refer to \cite{keiper2017compressed} for an introduction, as well as for a literature review on compressed sensing with binary prior and a sensing matrix that is Rademacher or Bernoulli.
		
		The two algorithms BIN-BP and P-BP empirically perform similarly in terms of required tests amount to succeed on the QGT problem.
		In the context of compressed sensing with general sparse signal, algorithms that require more running time have been developed to reduce the required number of tests.
		For instance, some algorithms iteratively solve convex optimization problems (see for example   \cite{candes2008enhancing,chartrand2008iteratively,dai2009subspace,wang2010sparse}).     
		These algorithms outperform BP in many cases, all of which consider non-binary signals. However, none of these more complicated algorithms that require more computation time perform much better than BIN-BP for the QGT problem. 
		See more details in \cite{wang2010sparse}. Specifically, see Figure 10.d in \cite{wang2010sparse} for simulations of these algorithms compared to BP on a binary signal.

		Algorithmic time complexity is the subject of other studies.
		An elementary and computationally very efficient algorithm is the Basic-Thresholding which we already mentioned in Section \ref{sec:contributions}.
		In most studies of Compressed Sensing the signal is not assumed to be binary. The $k$-Thresholding algorithm as we presented in Algorithm \ref{alg:Basic-Thresholding} is a variation to the special case of QGT.
		The Basic-Thresholding algorithm (as defined in \cite{foucart2013invitation} for example) uses 
		the $ k $-Thresholding (but with the $ \phi $ score function)
		to detect the \emph{support} of the signal, then it estimates the values of the entries in its support. 
		The support of a vector is the entries for which it is not zero.
		In the QGT problem, once we know a coordinate is in the support we also know that it is equal to one. 
		Hence, the variation we proposed for $k$-Thresholding.
		Adapting Basic-Thresholding to the QGT problem was already proposed in \cite{gebhard2019quantitative}, but with the original score function $ \{\phi_i\}_{i\in[n]} $.
		Similarly, \emph{Orthogonal Matching Pursuit} (OMP)\cite{pati1993orthogonal, davis1994adaptive} is a well known algorithm in the context of compressed sensing, and it was adapted to the finite or discrete signal prior \cite{sparrer2013discrete,sparrer2014adapting,sparrer2015soft}.
		The algorithms in these three papers perform similarly when noise is absent in the measurements.
		Therefore, as a representative we consider only the \emph{Q-OMP} \cite{sparrer2013discrete} algorithm in the empirical section (see test set \ref{simulations:testset:tbt-framework}  Section \ref{sec:empirical}). 

		The QGT problem also was studied under the assumption that $ k $ is of arbitrary-size (see \cite{soderberg1963combinatory,erdos1963two, cantor1964determining, lindstrom1965combinatorial, lindstrom1966combinatorial, lindstrom1971mobius, lindstrom1975determining}).
		In \cite{mangasarian2011probability} they show that for any $ k $ the system of linear equations $ Ax = y $ with minimization objective of the $ l_\infty $ norm has a single solution when $ m > (\nhalf+o(1))n $, i.e., solving this LP will yield exactly the defective items.

		Some works allow the tests to be planned ahead. 
		In such settings, algorithms have two phases: the \emph{test design phase}, in which the algorithm constructs the test matrix $ A $; and the \emph{the decode phase}, in which the algorithm needs to recover the defective items given $ y=Ax $.
		In \cite{lindstrom1975determining,djackov1975search} they prove that even when planned tests are allowed, at least $$ 2k\frac{\lognk}{\log k} $$ tests needs to be performed.
		In \cite{karimi2019sparse} and \cite{karimi2019non}, two closely related polynomial-time algorithms are presented for the sublinear regime and planned tests setting. 
		In both, an integer $ k $ (where $ k=O(n^\theta) $ for some $\theta\in \left(0,1\right)$) is given as a parameter before the test design phase. 
		In \cite{karimi2019sparse}, $k$ is the exact number of defective items and in \cite{karimi2019non}, each item is defective with probability $ \frac{k}{n} $. 
		The algorithms in \cite{karimi2019sparse} and \cite{karimi2019non}, using sparse graph codes over bipartite graphs, recover the defective items using about $ 1.19 k  \lognk $ and about $ 1.05 k \lognk $ tests, respectively.

		QGT was first introduced in \cite{shapiro1960e1399} by Shapiro in the \emph{adaptive setting}. 
		That is, the algorithm may plan tests iteratively, such that they are adapted to previous tests outcomes.
		The information theoretical lower bound on the number of tests in the adaptive setting is
		$$  
		(1-o(1)) \frac{k \lognk}{\log k}.
		$$ 
		Bshouty presented in \cite{bshouty2009optimal} an efficient algorithm for the adaptive QGT that performs
		$$  
		(1+o(1)) \frac{2k \lognk}{\log k}
		$$
		tests.

		As a more practical model, many Compressed Sensing works consider noisy measurements. 
		That is, they assume the input of the algorithm is $ y + \phi $ instead of $ y $ for some random vector $ \phi \in \mathbb{R}^m$. 
		This is a more realistic setting which has practical applications. 
		For simplicity, we don't consider noise in our model.
		However, the QGT problem with noise has some practical applications as in  bioinformatics \cite{cao2014quantitative} and traffic monitoring on the internet \cite{wang2015group}.
		We refer to \cite{foucart2013invitation,eldar2012compressed} and the references within for more information regarding applications and theoretical results of Compressed Sensing with noise.

	\subsection{Overview of techniques}\label{sec:tec_overview}
	
	Roughly speaking, we can divide the scope of this \paper\ into two parts. 
	The first is to prove which algorithms have the relaxed rank property. 
	The second is to show the required number of tests for an algorithm to succeed.
	
	We start with a proof sketch of Theorem \ref{thm:SSA-rank-condition}, see Section \ref{sec:proof:thm-relaxed-rank-is-enough} for a complete one.
	Consider an algorithm as in the theorem statement and denote its outcome with $ \algS $. 
	We have that $ \rankof{A|_\algS} > |\algS|-C\log n $ and that $ \D \subseteq \algS $ \whp.
	Assume both hold simultaneously.
	We have then that the system of linear equations $ (A|_\algS)z = y $ has at most $ C \log n $ free variables. 
	There are at most $ 2^{C\log n}=n^C $ different possible binary assignments to the free variables, so we can go over all of the possible binary solutions in polynomial time. 
	Additionally, as we know that $ \D \subseteq \algS $, there exists some binary solution $ x' \in \{0,1\}^{|\algS|} $ to this system of linear equations.
	Note that the required number of tests is above the information-theoretical threshold by assumption.
	Therefore, if a binary solution was found it is \whp indeed the defective items indicator vector $ x $.
	
	The main effort in this \paper\ is to prove that indeed Subset Select algorithms has the relaxed rank property. 
	We provide proof for some special cases in which that is the case.
	
	Known results regarding random matrix ranks are not sufficient to show that any Subset Select algorithm has the relaxed rank property.
	Consider an algorithm $ ALG $ for the Subset Select problem and denote its outcome with $ \algS $.
	Generally, to upper bound the probability that $ {\rankof{A|_S}} < m - C\log n $, one can use a union bound over all submatrices $ B \in  \{0,1\}^{m \times m}$ of a random Bernoulli matrix $ A \in \{0,1\}^{m \times n} $. 
	This gives us the upper bound of 
	\begin{equation}\label{discussion:eq:submatrix_union_bound}
	{n \choose m}\Pr\left[ \rankof{B} < m-C\log n \right]. 
	\end{equation}
	To the best of our knowledge, upper bounding the probability of a random square Bernoulli matrix to have such a rank was not studied. 
	The best known upper bound that we can use in this case is that of a square random Bernoulli matrix begin singular, that is the result $ \left(\frac{1}{2} + o(1)\right)^m $ due to a recent work \cite{tikhomirov2018singularity} after an extensive study of this bound in \cite{komlos1967determinant, kahn1995probability,tao2006random, tao2007singularity,rudelson2008invertibility,rudelson2008littlewood,tao2009inverse,bourgain2010singularity}.
	When considering the rank distribution of the matrix $ A|_\algS $ directly there are not many results either, as there are many dependencies among its entries for a non-trivial algorithm $ ALG $.
	It is not even a \emph{row} or \emph{column independent matrix} (see \cite{vershynin2010introduction} for details regarding this kind of random matrix results).
	
	In the following, we define some special cases of Subset Select algorithms for which we have a proof that they have the relaxed rank property.
	We define two properties, \emph{consistency} and \emph{second-stage-consistency}, for algorithms that solve the Subset Select problem.
	Intuitively, consider two problem instances that differ only in the content of a set $S\subseteq[n]\setminus\D$ of matrix columns that correspond to non-defective items.
	If $S$ is part of a consistent algorithm solution in both inputs, then the remaining part of the solution is also identical.
	In a second-stage-consistent algorithm, we require the above to hold only for some of the subsets $S\subseteq [n]\setminus\D$.
	Before we provide a more formal definition, we first introduce some notation.

	Consider a subset $ S \subseteq [n] $. 
	We define an equivalence class $ \sim_S $ on the binary matrix space $ \{0,1\}^{m \times n} $, such that two matrices are in the same equivalence class if they agree modulo the $ S $ columns. 
	Formally, for $ A,A' \in \{0,1\}^{m \times n} $ we say that $ A\sim_S A' $ if and only if $ A|_{\bar S} = A'|_{\bar S} $ (recall that $ \bar S $ is the complement of $ S $).
	We identify each element in the quotient set $ \{0,1\}^{m\times n} / \sim_S $ as $ [H]_{S} $ for $ H \in \{0,1\}^{m \times |\bar S|} $ by the canonical projection so that if $ A \in [H]_S $ then $ A|_{\bar S} = H $. 
	Finally, for an algorithm $ ALG $, denote its output on input test matrix $ A \in \{0,1\}^{m \times n} $ by $ \algS(A) $.
	
	\begin{de}\label{def:consistent}
		Fix $ \D $ and an algorithm $ ALG $.
		We say that $ ALG $ is \emph{consistent} if for every $ S \subseteq [n]\setminus \D $ and $ H \in \{0,1\}^{m\times |\bar{S}|} $ there exists a subset $ \algSH \subseteq [n] $ such that for each test matrix $ A\in [H]_{S} $, if $ S\subseteq \algS(A)$ then $ \algS(A)=\algSH $.
	\end{de}
	In Section \ref{sec:m-thresholding-is-consistent} we show that $m$-Thresholding is a consistent algorithm.
	The consistent property concerns subsets $ S\subseteq[n] $ for which $ S \cap \D = \emptyset $.
  The second-stage-consistent property concerns subsets $ S\subseteq[n] $ for which $ S \cap (\D \cup I(A)) = \emptyset $, where $ I(A) $ is the outcome of a QGT algorithm on input test matrix $ A \in \{0,1\}^{m \times n} $.
	In Section \ref{sec:second-stage-consistency-algs} we provide a property for QGT algorithms and prove that if an algorithm $ ALG $ satisfies it, then $ (ALG) $-Then-Thresholding is second-stage-consistent. 

  Formally, we define the second-stage-consistent property as follows.
  For a fixed $ \D $ and a Subset Select algorithm $ ALG $, we define a function $ I:=I_{\D,ALG} $	from the space of matrices $ \{0,1\}^{m \times n} $ to subsets of $ [n] $ such that
	\begin{equation}\label{SSA:eq:I_A-def}
		\forall A\in\{0,1\}^{m \times n} \;\text{ it holds that }\; I(A)\subseteq [n] \;\text{ and }\; |I(A)|\le k.
	\end{equation}
	\begin{de}\label{def:second-stage-consistent}
		Fix $ \D $ and an algorithm $ ALG $.
		We say that $ ALG $ is \emph{second-stage-consistent} if there exists a function $ I:=I_{\D,ALG} $ that satisfies (\ref{SSA:eq:I_A-def})  and for which the following holds: for every $ S \subseteq [n]\setminus \D $ and $ H \in \{0,1\}^{m\times |\bar{S}|} $ there exists a subset $ \algSH \subseteq [n] $ such that for every test matrix $ A\in [H]_{S} $, if $ S\subseteq \algS(A)$ and $ I(A)\cap S=\emptyset $ then $ \algS(A)=\algSH $.
		
		Additionally, we say that $ ALG $ is second-stage-consistent according to a function $ I $ if the above holds for $ I $.
	\end{de}		
	
	Finally, we define the threshold $ \hat{M}_{QGT}:=\hat{M}_{QGT}(n,k) \in \mathbb{N}$ such that for $ m> \hat{M}_{QGT}$ we have that
	\begin{equation*}
		{n-k \choose \min (k, m-k)}2^{-m} = o(1).
	\end{equation*} 	
	There is no known polynomial-time algorithm (to the best of our knowledge) that solves \whp the QGT problem using less than $ \hat{M}_{QGT} $ tests.
	Note that the threshold $ \hat{M}_{QGT} $ in the sublinear regime reads as $ \sim k \log \frac{n}{k} $. 
	
	\begin{lem}\label{SSA:lem:conditions}
		Consider an algorithm $ ALG $ that solves the Subset Select problem \whp for $ m>M $ where $ M=M(n,k) \in \mathbb{R} $. Denote $ ALG $'s outcome by $ \algS $. 
		There is a constant $ C\ge0 $ such that \whp the submatrix $ A|_\algS $ has rank of at least $ |\algS| - C \log n $ if any of the following conditions hold:
		\begin{enumerate}
			\item The outcome, $ \algS $, is of cardinality at most $ 2k $ and $ M \ge (1+\epsilon_n)\hMqgt $.
			\item The algorithm $ ALG $ is consistent.
			\item The algorithm $ALG$ is second-stage-consistent and $ M \ge (1+\epsilon_n)\hat{M}_{QGT} $.
		\end{enumerate}		
		(The term $ \epsilon_n = o(1) $ will be defined later.)
		
		Furthermore, the constant $ C $ in each of these cases equals $ 0,1 $ and $ 1 $, respectively.
	\end{lem}
	
	Roughly, the proof of the lemma (see Section \ref{sec:proof:lem-conditions}) goes as follows. 
	We show that for the majority of subsets $ S \subseteq [n] $ with cardinality $ \log n $, the entries of $ A|_S $ do not change much the outcome matrix $ A|_\algS $. Afterwards we apply a simple observation of Odlyzko.
	
	\begin{lem} \label{SSA:lem:X-in-V}
		\emph{\cite{odlyzko1988subspaces}} Let $ X $ be a random vector chosen uniformly at random from $ \{0,1\}^N $ and let $ V \subseteq \mathbb{R}^N $ be a linear subspace of dimension at most $ k $. Then,
		\begin{equation*}
		\Pr[X\in V] \le 2^{k-n}
		\end{equation*}
	\end{lem}		

	To analyze the effectivity of our proposed method, we give and compare thresholds for the $k$-Thresholding and $m$-Thresholding. 
	In the linear regime these algorithms need more than $ O(k \lognk) $ tests to succeed. Therefore, we analyze them only in the sublinear regime.
	
	\begin{lem}\label{boosting:lem:top-k-number-of-tests}
		Consider the sublinear regime, namely $ k=n^\theta $ where $ \theta \in (0,1) $ is a constant.
		Denote by $ \algS $ the outcome of $k$-Thresholding algorithm and $ M=M(n,k)\in \mathbb{R} $ as:
		\begin{equation}
			\label{boosting:lem:top-k-number-of-tests:eq}
			M := 2k  \frac{1 + \sqrt{\theta}}{1 - \sqrt{\theta}}  \ln \frac{n}{k}
		\end{equation}
		For $ m>M $ we have that $ \algS = \D $ \whp.		
	\end{lem}
	
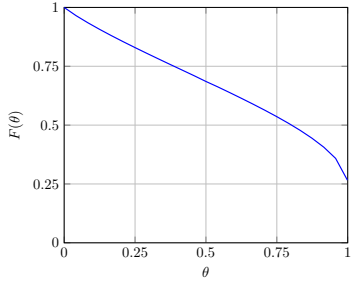
\begin{figure}
		\centering
\begin{tikzpicture}[scale=0.55]
\begin{axis}[
			xlabel={$ \theta$},
			ylabel={$ F(\theta) $},
			ytick={0,0.25,0.5,0.75,1},
			xtick={0,0.25,0.5,0.75,1},
			ymin=0,
			ymax=1,
			xmin=0,
			xmax=1,
			grid=both
			]
\addplot[domain=0:0.999, color=blue, style={thick}] {((2*x-1)/(x-1+sqrt(x-x^2)))^2 / ((1+sqrt(x))/(1-sqrt(x)))};

\end{axis}
\end{tikzpicture}

		\caption{
			The function $ F(\theta) $ is the ratio between Equation (\ref{boosting:lem:top-k-number-of-tests:eq}) and Equation  (\ref{boosting:lem:top-m-number-of-tests:eq}).
		}
		\label{figure:bt-ssbt-theory}
	\end{figure}

	\begin{lem}\label{boosting:lem:top-m-number-of-tests}
		Assume the sublinear regime.
		Denote by $ \algS $ the outcome of $m$-Thresholding algorithm and $ M=M(n,k)\in \mathbb{R} $ as: 
		\begin{equation}
			\label{boosting:lem:top-m-number-of-tests:eq}
			M:=2k\alpha^{2}\ln \frac{n}{k}, 
		\end{equation}
		where
		$$
		\alpha := 	\left\{
		\begin{array}{ll}
		2 & \mbox{if } \theta=\nhalf \\
		\frac{2\theta-1}{\theta-1+\sqrt{\theta(1-\theta)}} & \mbox{otherwise. }
		\end{array}
		\right.
		$$
		For $ m>M $ we have that $  \D \subseteq \algS $ \whp.
	\end{lem}	
	Let $F(\theta)$ denote the ratio between the threshold given in Lemma \ref{boosting:lem:top-k-number-of-tests} for $ k $-Thresholding and the threshold given in Lemma \ref{boosting:lem:top-m-number-of-tests}
	for $ m $-Thresholding, for a given value of $\theta$. For the reader’s convenience, $F(\theta)$ is plotted in Figure \ref{figure:bt-ssbt-theory}. 
	As $\theta$ approaches~$ 1 $, $F(\theta)$ approaches $\frac{1}{4}$, showing that the threshold given in Lemma~\ref{boosting:lem:top-m-number-of-tests}
	for $ m $-Thresholding is up to $ 4 $ times better than that given in Lemma~\ref{boosting:lem:top-k-number-of-tests} for $ k $-Thresholding. When $\theta$ approaches~$ 0 $, $F(\theta)$ approaches~$ 1 $, and the
	two thresholds converge to each other.
	
	In test set \ref{simulations:testset:bt-ssbt-2kssbt} (Section \ref{sec:empirical}) we performed simulations and measured the success frequency of the $ k $-Thresholding and the $ m $-Thresholding algorithms. 
	The	success frequencies observed in the simulations agree with the theoretical analysis in Lemma \ref{boosting:lem:top-k-number-of-tests} and Lemma \ref{boosting:lem:top-m-number-of-tests}.

	\subsection{Discussion}\label{sec:discussion}
	
		In this {\paper} we present the Subset Select problem and show a way how to apply Subset Select algorithms to the QGT problem.
		We provide a framework to convert QGT algorithms to the Subset Select problem in order to reduce the number of required tests.
		However, one may consider different ways to use the Subset Select problem to induce QGT algorithms. 
		For example, consider two QGT algorithms.		
		Suppose that $ m > 2k $, and define the Subset Select algorithm that given a problem instance, it runs these two QGT algorithms and combines their outcomes.
		This Subset Select algorithm succeeds not only if one of these QGT algorithms succeeds, but also if their combined outcomes covers $ \D $.
		By Lemma \ref{SSA:lem:conditions}, because this Subset Select algorithm outcome is of cardinality $ 2k $, it has the relaxed rank property. 
		Hence, it can be used to solve the QGT problem.
		We find it interesting to study new algorithms using the Subset Select problem.
		
		In the context of the QGT problem, the Subset Select problem raises the following questions which we find interesting.
		
		\begin{question}
			In our simulations, the outcome $ \algS $ of $m$-Thresholding was always such that $ \rankof{A|_\algS} = m $.
			We prove in Lemma \ref{boosting:lem:basic-thresholding-is-consistent} that $ \rankof{A|_\algS} \ge m - \log n$ \whp. 
			Does $ \rankof{A|_\algS} = m $ hold \whp as the simulations suggests? 
		\end{question}

		\begin{question} \label{question:existence_of_singular_correct_answer}
			Consider the integers $ M:=M(N) \in \mathbb{N}$ and $ K:=K(N) \in \mathbb{N}$ such that $ 0\le K\le M\le N $, and a random Bernoulli matrix $ A \in \{0,1\}^{M \times N} $.
			What is the probability that there exists a subset $ S \subseteq [n] $ of cardinality $ M $ such that $ \{1,2,\ldots, K\} \subseteq S $ and $ \rankof{A|_S} < M $?		
		\end{question}
		Suppose the answer to Question \ref{question:existence_of_singular_correct_answer} is bounded by $ o(1) $ for some parameter range (i.e., for some functions $ M $ and $K $).
		Note that the answer remain the same also if we change the question by replacing the set $ \{1,2,\ldots,K\} $ with some arbitrary fixed subset $ \D\subseteq [n] $ of cardinality $ K $.
		Therefore, in this parameter range and $ m=M, k=K $, a Subset Select algorithm that satisfies	\whp the first condition of Theorem \ref{thm:SSA-rank-condition} (i.e., the subset $ \D $ is contained in its outcome) also has the relaxed rank property \whp.
		Namely, such answer implies that the second condition in Theorem \ref{thm:SSA-rank-condition} is redundant for this parameter range.
		
		\begin{question}\label{question:rank-bound}
			Consider a random Bernoulli matrix $ B \in \{0,1\}^{M \times M} $.
			Does the following upper bound hold?
			$$
			Pr\left[ \rankof{B} \le R \right]  
			\leq \left (1+o(1)\right )\Pr\left[ B \text{ is singular} \right]^{M-R} 			
			$$
		\end{question}
		
		If the answer to Question \ref{question:rank-bound}  is positive, then together with Equation (\ref{discussion:eq:submatrix_union_bound}) and the result of \cite{tikhomirov2018singularity} that 
		$
		\Pr\left[ B \text{ is singular} \right] \le (\nhalf + o(1))^{M}
		$
		we have that \whp $ \rankof{A|_S} \ge m-O(\log n)$ for all subsets $ S\subseteq [n] $ of cardinality $ m $.
		
		\begin{question}\label{question:algo}
			Suppose that the answer to Question \ref{question:rank-bound} is negative.
			Consider the integer $ M:=M(N) \in \mathbb{N}$ such that $ 0\le M\le N $ and a random Bernoulli matrix $ A \in \{0,1\}^{M \times N} $.
			Does there exist a polynomial-time algorithm that finds a subset $ S $ such that $ |S| \le M $ and $ \rankof{A|_S} \le |S| -O(\log N)$?
		\end{question}
			Consider some parameter range such that the answer to Question \ref{question:algo} is negative.
			In this case, any polynomial-time algorithm that solves the Subset Select problem will have the relaxed rank property.
			The problem of finding a small set of linearly dependent vectors was considered in \cite{bhattacharyya2011complexity}. An exponential time algorithm has been proposed for a similar, although not random, setting.

	\section{Proofs}
	
	\subsection{Proof of Theorem \ref{thm:SSA-rank-condition}}
	\label{sec:proof:thm-relaxed-rank-is-enough}
	\begin{proof}
		Consider a Subset Select algorithm $ ALG $ that satisfies the theorem conditions. We use the notations $ \algS, C$ and $ M $ as in the theorem statement.
		We now define an algorithm $ ALG' $ for the QGT problem.
		Given a problem instance $ (A,y,k) $, $ ALG' $ returns the outcome of  Recover-From-Submatrix (see Algorithm \ref{alg:recover-D-from-S}) on the problem instance $ (A,y,k) $ together with $ S:=\algS $ (the outcome of $ ALG $ on this problem instance). 
		\begin{algorithm}[!htbp]
			\caption{Recover-From-Submatrix}
			\label{alg:recover-D-from-S}
			\begin{algorithmic}
				\STATE \textbf{Input:} QGT problem instance $ (A,y,k) $ and a subset $ S = \left\{s_1, s_2, \ldots, s_{|S|}\right\} $
				\begin{enumerate}
					\item Using Gaussian elimination, compute the {reduced-row-echelon\footnotemark} form of the system of linear equations $ \left(A|_S\right)z'=y $ and denote it by $ Bz'= v $
					\item Denote by $ F \subseteq [n] $ the indices of all {free-variables\footnotemark} 
					in $ Bz'=v $ 		
					\item For every subset $ P \subseteq F $ do: \label{alg:line:for-loop}
					\begin{enumerate}
						\item Check if the (unique) solution of following is binary:
						\label{alg:line:consistent-check}
						\begin{equation*}
						\begin{split}
						Bz'=v \\
						z'_i=1 &  \text{  }\forall i \in P \\
						z'_i=0 &  \text{  }\forall i \in F \setminus P \\
						\end{split}
						\end{equation*}
						\item If $ z' \in \{0,1\}^{|S|} $ and of Hamming weight $ k $, return $ \algS := \{s_i : z'_i = 1\} $
					\end{enumerate}
				\end{enumerate}
			\end{algorithmic}
		\end{algorithm}%
			\addtocounter{footnote}{-1} 
	\footnotetext{
		A matrix is in \emph{reduced row echelon} form if: all rows consisting of only zeroes are at the bottom; the leading coefficient of a nonzero row equals $ 1 $ and is strictly to the right of the leading coefficient of the row above it; the column containing this leading $ 1 $ coefficient has zeros in all its other entries. 						
	}
	\addtocounter{footnote}{+1} 
	\footnotetext{
		The columns that don't correspond to any row with a leading $ 1 $ coefficient are \emph{free variables}. 
	}
		
		In the following we will show that $ ALG' $ is an algorithm for the QGT problem which fulfill the theorem statement.
		Recall that $ \Mqgt $ is the information theoretic threshold defined in the beginning of Section \ref{sec:introduction} and assume that $ m>\max(M,M_{QGT}) $. 
		For a QGT problem instance $ (A,y,k) $, we will show that if \whp $ \D\subseteq \algS $ and $ \rankof{A|_{\algS}} > |\algS|-C\log n $ then \whp $ ALG' $ recovers $ \D $ in polynomial-time.

		By the definition of $ M_{QGT} $, \whp there exists a unique solution $ z \in \{0,1\}^n $ of Hamming weight $ k $ for the system of linear equations $ Az = y $. 
		Namely, if such a vector $ z $ is a solution for $ Az = y $, then \whp it equals $ x $.
		When Algorithm \ref{alg:recover-D-from-S} does not fail, it outputs a subset $ S \subseteq [n] $.
		Observe that $ S $ is such that $ \one_{S} \in \{0,1\}^n $ is a solution of Hamming weight $ k $ for this system of linear equations.
		Therefore, if the algorithm returns a solution then it is \whp correct.
		Assume that $ \D \subseteq \algS $, then there is a binary solution to the system of linear equations $ Bz'=v $ of Hamming weight $ k $.
		Note that the algorithm goes over all possibly binary solutions of $ Bz' = y $.
		Therefore, if \whp $ \D \subseteq \algS $ and $ \rankof{A|_\algS}>|\algS|-C\log n $, then the algorithm $ ALG' $ recovers $ \D $ \whp.

		We are left to show that the algorithm returns a solution within the required time complexity.
		Recall that Gaussian elimination algorithm time complexity is $ O(m^3) $.
		Because the rank of $ A|_S $ is not smaller than $ |S| - C\log n $ there are at most $ C\log n $ free variables in $ Bz' = v $, i.e., $ |F| \leq C \log n $.
		Hence, the exhaustive search (in line \ref{alg:line:for-loop}) goes over $ 2^{|F|}\le n^C $ different systems of linear equations.
		Because the matrix $ B $ is of a row-echelon form with $ m $ equations and at most $ C\log n $ free variables, it takes to check the condition at line \ref{alg:line:consistent-check} only $ O(m\log n) $ operations. Hence, we get the required time complexity.	
	\end{proof}
	
	\begin{remark}
		We can replace $ {O}(m^3 +n^c m \log n) $ with $ \tilde{O}(n^c T(m)) $, where $ T(m) $ is the computational complexity of solving a system of linear equations with m linear equations and m unknowns.
	\end{remark}

	\subsection{Proof of Lemma \ref{SSA:lem:conditions}}
	\label{sec:proof:lem-conditions}
	\begin{proof}
		Fix $ \D $. We will prove separately, that if either of the conditions stated in Lemma \ref{SSA:lem:conditions} holds, the submatrix $ A|_\algS $ has rank of at most $ |\algS| - C \log n $.
		
		\emph{Condition 1}. 
		Suppose that the first condition holds. 
		Recall that by the definition of the Subset Select problem we have that $ |\algS| \le m $ and consider a non-defective items subset $ S \subseteq \notD $ of cardinality $ \min (k, m-k) $.
		Denote by $ L_{S} $ the event that the columns of $ \D \cup S $ in the test matrix are linearly dependent (i.e., the event that $ \rankof {A|_{\D\cup S}} < |S\cup \D| $).
		By \cite{tikhomirov2018singularity}, the probability of a random Bernoulli matrix $ B\in\{0,1\}^{m\times m} $ to be singular is bounded above by $ (\nhalf + o(1))^m $.
		Therefore and because $ |S\cup\D|\le m $, we have that:
		\begin{equation}\label{SSA:lem-proof:eq:L_S-bound}
			\Pr[L_S] = \Pr[\rankof{A|_{S\cup \D}<|S\cup \D|}] \le \left (\nhalf+o(1)\right )^{-m}.
		\end{equation}
		By using a union bound, the probability that one of the bad events in $ \{L_S \;|\; S \subseteq \notD \;, |S|= \min (k, m-k)  \} $ did occur is bounded by
		\begin{equation*}
			\Pr \left[\bigvee_S L_S \right] 
			\le {n-k \choose \min (k, m-k)} \left (\nhalf+o(1)\right )^{-m}.
		\end{equation*}
		By the definition of $ \hat{M}_{QGT} $, there exists some $ \epsilon_n  \rightarrow 0$ such that for $ m \geq  (1+\epsilon_n)\hat{M}_{QGT}$ we have 
		\begin{equation}
			\label{SSA:lem-proof:eq:union-L_S-bound}
			\Pr \left[\bigvee_S L_S \right] 
			\le o(1).
		\end{equation}
		Hence, for every subset $ S $ with $ \D \subseteq S $ and $ |S| \le 2k $, we have that \whp $ {\rankof{A|_{S}}} = |S| $. 
		
		By the Condition 1 assumption $ ALG $ solves \whp the Subset-Select problem for $ m>M \ge (1+\epsilon_n)\hat{M}_{QGT}  $ and its outcome $ \algS $ is of cardinality at most $ 2k $. 
		Therefore, we conclude that the theorem statement holds for  Condition 1 with $ C=0 $.
	
		\emph{Condition 2}. Assume that the second condition holds.		
			In the following we show that for $ 1 \le t  \le m-k$ the probability that $ A|_{\algS} $ is of rank less than $ m -t $ is bounded above by $ {n \choose t}2^{-t^2} +o(1)$. By assigning $ t=\log n $ the proof will follow.

			Denote by $ B $ the bad event $ \rankof{A|_{\algS}} < m -t $
			and, for this proof only, we use the notation $ A_S $ to describe the following set of vectors
			$$ A_S := \{A_i | i\in S\}, $$ 
			where $ A_i $ is the $ i $th column.
			It is then that $ \langle A_S \rangle $ is the vector space spanned by the columns of $ S $ in the test matrix $ A $.
			Note that when $ B $ occurs there are $ t $ columns in $ A|_{\algS} $ that are linearly dependent on the other columns. 
			Formally, there exist a subset $ T\subseteq [n] $ of cardinality $ t $ such that 
			\begin{equation}\label{SSA:lem-2:eq:B_S-condition}
				T\subseteq\algS \;\text{ and }\; \langle A_T \rangle \subseteq \langle A_{\algS \setminus T} \rangle.
			\end{equation}
			
			Towards a union bound argument, we provide a cover of the event $ B $. 
			For every subset $ T \subseteq [n] $ of cardinality $ t $, denote by $ B_T $ the event that (\ref{SSA:lem-2:eq:B_S-condition}) has happened.
			From the above, note that
			\begin{equation*}
			\Pr[B] = \Pr\left[\bigvee_T B_T\right].
			\end{equation*}
			
			We now give a brief overview of the rest of this proof.
			First, we get rid of each event $ B_T $ for which $ T \cap \D \neq \emptyset $ by showing that, \whp, if $ B_T $ occurs for some $ T \cap \D \neq \emptyset $, then there exists a $ B_{T'} $ with $ T' \cap \D = \emptyset $ that also occurs.
			We then only need to bound events $ B_T $ with $ T \cap \D = \emptyset $.
			Consider such subset $ T \subseteq \notD $ and an assignment of columns other than $ T $.
			From the consistency of $ ALG $, there exists a fixed subset $ S'\subset [n] $ such that if $ T \subseteq \algS $ then $ \algS = S' $. 
			In other words, when columns in $ \bar{T} $ are fixed, the event $ B_T $ (which implies $ T \subseteq \algS $) occurs  when $ \langle A_T \rangle \subseteq \langle A_{S' \setminus T} \rangle $, where $ \langle A_{S' \setminus T} \rangle $ is a deterministic subspace.				
			Therefore, we may use lemma \ref{SSA:lem:X-in-V} to bound with $ 2^{-t^2} $ the probability that $ t $ i.i.d.\@ random binary vectors are in a deterministic subspace of dimension at most $ m-t$. 
			Because there are no more than $ {n-k \choose t} $ such subsets we get the required bound. 			
		
			Denote by $ L_\emptyset $ the event that the columns of $ \D $ are linearly dependent and by $ \bar{L}_\emptyset $ its complementary.
			From inequality (\ref{SSA:lem-proof:eq:L_S-bound}) it follows that $ \Pr[L_\emptyset] \le o(1) $. 
			Therefore,
			\begin{equation*}
			\begin{split}
			\Pr[B] 
			&= \Pr\left[\bigvee_T B_T\right] \\
			&\le \Pr[L_\emptyset] + \Pr\left[\bigvee_T B_T \prmid \bar{L}_\emptyset\right] \\
			&\le o(1) + \Pr\left[\bigvee_T B_T  \prmid \bar{L}_\emptyset\right].
			\end{split}
			\end{equation*}
			
			Conditioned on that $ \bar{L}_\emptyset $ has happened, suppose that $ B_T $ occurred for some $ T \subseteq [n] $ with $ T \cap \D \neq \emptyset $.
			In this case, the rank of $ A|_\algS $ is lower that $ |\algS|-t $.
			Because the columns of $ \D $ are linearly independent, there exists a basis $ \mathcal{B}\subseteq A_\algS $ of the vector space $ \langle A_\algS \rangle $ such that $ \D\cap\algS \subseteq \mathcal{B} $.
			Therefore, there exists a subset $ T' \subseteq \algS $ such that $ B_{T'} $ occurs and $ T' \cap \D = \emptyset $.			
			Hence,
			\begin{equation*}
			\Pr\left[\bigvee_T B_T \prmid \bar{L}_\emptyset\right] = \Pr\left[\bigvee_{T:T\cap \D = \emptyset} B_T \prmid\bar{L}_\emptyset\right].
			\end{equation*}
			Using a union bound we get
			\begin{equation*}
			\Pr\left[\bigvee_{T:T\cap \D = \emptyset} B_T \prmid\bar{L}_\emptyset\right]
			\leq \sum_{T: T\cap \D = \emptyset} \Pr\left[ B_T \prmid\bar{L}_\emptyset\right].
			\end{equation*}
			The sum goes over exactly $ n-k \choose t $ subsets, so we are left to show for a fixed subset $ T \subseteq [n] \setminus \D $ of cardinality $ t $ that $ \Pr[B_T | \bar{L}_\emptyset] \le 2^{-t^2}$.
			
			By the law of total probability we have that
			\begin{equation*}
			\begin{split}
			\Pr\left [B_T \prmid \bar{L}_\emptyset\right ]
			&= \sum_{H \in \{0,1\}^{m\times (n-t)}} 
						\Pr\left[A|_{\bar{T}}=H \prmid \bar{L}_\emptyset \right] 
						\Pr\left [B_T \prmid A|_{\bar{T}}=H\right].
			\end{split}
			\end{equation*}		
			Consider a subset $ T \subseteq \notD $ of cardinality $ t $ and $ H \in \{0,1\}^{m \times (n-t)} $, we will show that 
			$$
				\Pr\left[B_T \prmid { A|_{\bar{T}}=H}\right] \le 2^{-t^2}.
			$$
			The proof of Condition 2 of Lemma \ref{SSA:lem:conditions} will follow from 
			\begin{equation*}
			\Pr\left [B_T \prmid \bar{L}_\emptyset\right ] 
			\le \sum_{H \in \{0,1\}^{m\times (n-t)}} \Pr\left [A|_{\bar{T}}=H\prmid\bar{L}_\emptyset\right ] 2^{-t^2}
			= 2^{-t^2}.
			\end{equation*}
			
			Now, use the explicit notation $ \algS(A) = \algS $ for the outcome of $ ALG $ on a test matrix $ A $.
			From the consistency of $ ALG $ and because $ T \cap \D = \emptyset $, there exists a subset $ \algSTH \subseteq [n] $ such that for all $ A \in [H]_T $ if $ T \subseteq \algS(A) $, then $ \algS(A) = \algSTH $.			
			Denote by $V_{H,T} $ the subspace spanned by the columns of $ \algSTH \setminus T $ (i.e., $ V_{H,T} = \langle A_{\algSTH \setminus T} \rangle $). 
			Note that because $ (\algSTH \setminus T) \subseteq \bar{T} $ the subspace $ V_{H,T} $ is deterministic.
			Hence, together with the definition of $ B_T $ we have that
			\begin{equation*}
			\begin{split}
			\Pr\left [B_T \prmid  A|_{\bar{T}}=H\right ] 
			&= \Pr\left [T\subseteq \algS \land \langle A_T\rangle \subseteq V_{H,T} \prmid  A|_{\bar{T}}=H\right ] \\
			&\le \Pr\left [\langle A_T\rangle \subseteq V_{H,T} \prmid  A|_{\bar{T}}=H\right ].
			\end{split}
			\end{equation*}
			Furthermore, because $ A_T $  is independent of $ A|_{\bar{T}} $ and $ T\cap \D = \emptyset $ we have:
			\begin{equation}\label{SSA:lem-cond2:eq:final}
				\Pr\left [\langle A_T\rangle \subseteq V_{H,T} \prmid  A|_{\bar{T}}=H\right ]
			= \Pr\left [\langle A_T\rangle \subseteq V_{H,T} \right ].
			\end{equation}
			Now, because $ V_{H,T} $ is spanned by less than $ m-t $ vectors its dimension is at most $ m-t-1 $. 
			Finally, bound (\ref{SSA:lem-cond2:eq:final}) with the probability of $ t $  i.i.d.\@ random binary vectors to belong the deterministic subspace $ V_{H,T} \subseteq \mathbb{R}^m$. From Lemma \ref{SSA:lem:X-in-V}, we have that 
			\begin{equation*}
			\Pr\left [\langle A_T\rangle \subseteq V_{H,T} \right ]\le 2^{-t^2}.
			\end{equation*}

		\newcommand{\Gevent}{\bar{L}}
		\emph{Condition 3}.
		This proof is similar to the one of condition 2.
		Assume that the third condition holds and that $ m > 2k $, otherwise the lemma holds by Condition 1.
		Let $ I:=I_{\D, ALG} $ be the function for which $ ALG $ is second-stage-consistent.		
		Denote by $ L := L_{I(A)} $ the event that the columns of $ I(A) \cup \D $ are linearly dependent.
		Note that $ L $ is covered by the events in $ \{L_S | S \subseteq \notD \;, |S|= k \} $ (as defined in Condition 1 proof).
		Therefore, from (\ref{SSA:lem-proof:eq:union-L_S-bound}) together with the assumption that $  M > (1+\epsilon_n) \hat{M}_{QGT}$, we have that 
		\begin{equation*}
			\Pr \left[ L \right]
			\le o(1).
		\end{equation*}			
		As before, consider the event $ B $ and its cover $ \{B_T: T\subseteq [n], \; |S|=t\} $. 
		Now, similarly, we bound $ B $ as follows:
		\begin{equation*}
			\Pr\left[B\right] 
			\le o(1) + \Pr\left[\bigvee_T B_T \prmid \Gevent\right]. 			 
		\end{equation*}
		Conditioned that $ {\Gevent} $ happened, whenever $ B_T $ occurs for some subset $ T $ with $ T \cap (\D \cup \SA) \neq \emptyset $, there exists a subset $ T' $ with $ T' \cap (\D \cup \SA) = \emptyset $ such that $ B_{T'} $ also occurs.		
		Denote by $ D_T $ the event that $ T \cap \SA = \emptyset $.
		To conclude
		\begin{equation*}
		\Pr\left[\bigvee_T B_T \prmid \Gevent\right] 
		= \Pr\left[\bigvee_{T:T\cap \D = \emptyset} B_T\land D_T \prmid\Gevent\right].
		\end{equation*}
		By using a union bound we get
		\begin{equation}
		\Pr\left[\bigvee_{T:T\cap \D = \emptyset} B_T \land D_T \prmid\Gevent\right]
		\leq \sum_{T: T\cap \D = \emptyset} \Pr\left[ B_T \land D_T \prmid\Gevent\right].
		\end{equation}
		
		To finish this proof, while not repeating ourselves, we only show that for any relevant subset $ S\subseteq \notD $ and $ H \in \{0,1\}^{m \times n-t} $ the following holds
		$$\Pr\left [B_T \land D_T \prmid \Gevent, A|_{\bar{T}}=H\right ] \le 2^{-t^2}.$$ 
		
		Now, use the explicit notation $ \algS(A) = \algS $ for the outcome of $ ALG $ on a test matrix $ A $.
		Recall that $ ALG $ is second-stage-consistent. 
		Hence, there exists a deterministic subspace $  V_{H,T} $ such that whenever $ D_T $ occurs and $ T \subseteq  \algS(A) $ holds, we have that $ \langle A_{\algS(A) \setminus T} \rangle =  V_{H,T}$. 
		We conclude that
		\begin{equation*}
		\begin{split}
		\Pr\left [B_T  \land D_T \prmid \Gevent, A|_{\bar{T}}=H\right ] 
		&= \Pr\left [T\subseteq \algS(A) \land \langle A_T\rangle \subseteq  V_{H,T} \land D_T \prmid \Gevent, A|_{\bar{T}}=H\right ] \\
		&\le \Pr\left [\langle A_T\rangle \subseteq  V_{H,T} \land D_T \prmid \Gevent, A|_{\bar{T}}=H\right ] \\ 
		& = \Pr\left [\langle A_T\rangle \subseteq  V_{H,T} \land D_T \right ] \\ 
		&\le\Pr\left [\langle A_T\rangle \subseteq  V_{H,T} \right ] \\
		&\le 2^{-t^2}.		
		\end{split}
		\end{equation*}
		And we done.
	\end{proof}

	\subsection{$m$-Thresholding is consistent}
	\label{sec:m-thresholding-is-consistent}
	\begin{lem} \label{boosting:lem:basic-thresholding-is-consistent}
		$m$-Thresholding is consistent, and hence it has the relaxed rank property.
	\end{lem}
	\begin{proof}
		Fix $ \D $ and denote by $ \algS(A) $ the outcome of $m$-Thresholding on a test matrix $ A\in\{0,1\}^{m \times n} $. 
		Consider a subset $ S\subseteq [n]\setminus\D $ and a matrix $ H \in \{0,1\}^{m\times |\bar{S}|} $.
		We need to provide a subset $ \algSH $ such that for all $ A\in [H]_{S} $, if $ S \subseteq \algS(A)  $ then $ \algS(A) = \algSH $.
		Assume $ |S|\le m $, otherwise the claim trivially holds.
		
		Recall that the score $ \psi_i $ of item $ i\in[n] $ is a function of the $ i $th test matrix column and of the tests outcome $ y $.
		Because $ S \cap \D = \emptyset $, for every $ A,A' \in [H]_{S} $ we have that $ Ax=A'x $.
		Namely, the tests outcome vector is equal for all tests matrices in $ [H]_{S} $.
		Therefore, the score $ \psi_i $ of an item $ i\in \bar{S} $ has fixed value for all tests matrices in $ [H]_{S} $.
		Sort the items in $ \bar{S} $ by descending order of score.
		Denote by $ S' \subseteq \bar{S} $ the subset that contains the first $ m-|S| $ items.		
		Note that the subset $ S' $ is	determent by $ H $ and $S$. 
		Define the subset $ \algSH := S'\cup S $. 
		We have that if $ S \subseteq \algS(A)$ then $ \algS(A)=\algSH $.
	\end{proof}

	\subsection{Sufficient condition for second-stage-consistency}
	\label{sec:second-stage-consistency-algs}
		Consider a QGT algorithm $ ALG $ and the Subset Select algorithm produced by our proposed framework. 
		That is, the $ (ALG) $-Then-Thresholding algorithm.
		We now discuss the reason why $ (ALG) $-Then-Thresholding may not be consistent.
		Let  $ S^\dagger $ and $ \algS $ denote the outcomes of the QGT algorithm $ ALG $ and of the Subset Select algorithm ($ ALG $)-Then-Thresholding, respectively.
		Suppose that for some problem instance, $ ALG $ did not recover all defective items (i.e, $ S^\dagger  \neq \D $).
		Recall, that the $ (ALG) $-Then-Thresholding algorithm calculates the score $ \phi_i $ for an item $ i $, which depends on the test matrix columns of $ S^\dagger $, $ \D $ and $ i $. 
		Then, ($ ALG $)-Then-Thresholding returns $ S^\dagger $ together with the first $ m-k $ items sorted by score.
		Note that these score values depend also on the non-defective items that were in $ S^\dagger $, denote these non-defective items by $ I\subseteq S^\dagger\setminus\D $.
		The column entries of items in $ I $ may affect the algorithm outcome $ \algS $ in an inconsistent manner.
		That is, there might be two test matrices that differ only in the contents of the $ I $ columns such that for both of these test matrices, all of the items in $ I $ are part of the algorithm's outcome $ \algS $, however, the rest of the outcome is not identical.
		In this case, ($ ALG $)-Then-Thresholding is not consistent.
		Nevertheless, ($ ALG $)-Then-Thresholding may be second-stage-consistent (for instance if $ ALG $ satisfies the condition in Lemma \ref{boosting:lem:when-is-second-stage-consistent}).
		Intuitively, a second-stage-consistent algorithm has at most $ k $ non-defective items whose column entries affect the algorithm outcome in an inconsistent manner.
		As we formally define in Definition \ref{def:second-stage-consistent}, 
		these $ k $ non-defective items may be determined after observing $ A $.
		
		Consider an algorithm $ ALG $ for the QGT problem and
		denote its output on an input test matrix $ A \in \{0,1\}^{m \times n} $ by $ S^\dagger(A) $.		
		In Lemma \ref{boosting:lem:when-is-second-stage-consistent} we define a property such that if $ ALG $ satisfies it, then $ (ALG) $-Then-Thresholding is second-stage-consistent.
		Intuitively the property is as follows.
		Consider two problem instances that differ only in the content of a set $S\subseteq[n]\setminus\D$ of matrix columns that correspond to non-defective items.
		If in each of the inputs, $ S $ is not part of $ ALG $'s solution, then the two solutions are identical.

		\begin{lem} \label{boosting:lem:when-is-second-stage-consistent} 						
			If for each subset $ S \subseteq \notD $ and matrix $ H \in \{0,1\}^{m \times |\bar{S}|} $ there exists a subset $ S^\dagger_{H,S} $ such that for all $ A\in \{A'\in[H]_{S} \;|\; S \cap S^\dagger(A') = \emptyset \}$ it holds that $ S^\dagger(A) = S^\dagger_{H,S} $, then the Subset Select algorithm $ (ALG) $-Then-Thresholding is second-stage-consistent with the function $ I(A) := S^{\dagger}(A) $.
			
			Hence, if $ ALG $ satisfies the above, $ (ALG)-Then-Thresholding $ has the relaxed rank property.
		\end{lem}
		Before we prove the above, note that $k$-Thresholding algorithm satisfies the requirements of Lemma \ref{boosting:lem:when-is-second-stage-consistent}.
		Therefore, 
		\begin{lem}
			The ($k$-Thresholding)-Then-Thresholding algorithm is second-stage-consistent.
		\end{lem}
		Now, consider the algorithm \emph{Iterative-Thresholding} (see Algorithm \ref{alg:Iterative-Thresholding} in Section \ref{sec:empirical}). Starting with some subset $ \algS = \emptyset $, this algorithm calculates for each item $ i $ its score $ \phi_i = \langle A, y-A\one_\algS \rangle / \|A_i\|_1$.
		Iteratively it adds to $ \algS $ the item with the highest score $ \phi_i $ (excluding items that are already in $ \algS $) and updates all scores according to the new $ \algS $. 
		The algorithm halts and returns $ \algS $ after $ k $ iterations (i.e., when $ |\algS| = k $).
		Note that this algorithm also satisfies the requirement of Lemma \ref{boosting:lem:when-is-second-stage-consistent} and therefore
		\begin{lem}
			The (Iterative-Thresholding)-Then-Thresholding algorithm is second-stage-consistent.
		\end{lem}
	
		\begin{proof}[Proof of Lemma \ref{boosting:lem:when-is-second-stage-consistent}]
			Denote by $ \algS(A) $ the outcome of $ (ALG) $-Then-Thresholding on the test matrix $ A\in \{0,1\}^{m \times n} $. 
			Consider a subset $ S\subseteq \notD $, a matrix $ H \in \{0,1\}^{m\times |\bar{S}|} $ and the function $ S^{\dagger}(A) $.	
			We need to provide a subset $ \algSH $ such that for all  $ A \in [H]_{S} $ if $ S \subseteq \algS(A)  $ and $ S \cap S^\dagger(A) = \emptyset $ hold then $ \algS(A) = \algSH $.
			Namely, $ \algS(A)=\algSH$ for each test matrix $ A\in[H]_S $ such that $ S $ is part of the new outcome (i.e., $ \algS(A) $), but was not part of the original outcome (i.e., $ S^\dagger(A) $).
			Denote the set of relevant matrices as
			$$ 
			\mathcal{H}=\mathcal{H}_{H,S,ALG} := \left \{A \in [H]_S \prmid S^\dagger(A)\cap S = \emptyset \,\land\,  S \subseteq \algS(A)\right \}. 
			$$			
			Assume $ |S|\le m-k $, otherwise the claim trivially holds (where the minus $ k $ is because $ S^\dagger(A) \subseteq \algS(A) $ by the definition of ($ ALG $)-Then-Thresholding).			
			Observe that by the definition of $ (ALG) $-Then-Thresholding the score $ \phi_i $ of an item $ i \in [n] $ depends only on the test matrix columns of $ S^\dagger(A) \cup \D \cup \{i\} $.
			Consider the subset $ S^\dagger_{H,S} $ as defined in the lemma statement.
			By definition, for all $ A\in \mathcal{H} $ because $ S\cap S^\dagger= \emptyset$ we have that $ S^\dagger=S^\dagger_{H,S} $.
			Therefore, the score $ \phi_i $ for $ i\in \bar{S} $	is fixed for all test matrices in $\mathcal{H}$.	
			Sort the items in $ \bar{S} $ by descending order of this fixed scores.
			Denote by $ S' \subseteq \bar{S} $ the subset that contains the first $ m-k-|S| $ items.
			Define the subset $ \algSH := S'\cup S \cup S^\dagger_{H,S}$. 
			The subset $ S'\cup S^\dagger_{H,S}$  is fixed for any test matrix $ A\in\mathcal{H} $ and additionally $ S'\cup S^\dagger_{H,S} \subseteq \algS(A)$. 
			Therefore, for every $ A\in \mathcal{H} $ if $ S \subseteq \algS(A)$ then $ \algS(A)=\algSH $.					
		\end{proof}

	\subsection{Proofs of Lemma \ref{boosting:lem:top-k-number-of-tests} and Lemma \ref{boosting:lem:top-m-number-of-tests}} \label{boosting:subsec:basic-thresholding}
	Recall that $ k $-Thresholding is a variation of the Basic-Thresholding algorithm and that the score for item $ i\in[n] $ in the Basic-Thresholding algorithm defined as $ \phi_i:=\sum_{j=1}^m A_{ji}y_j / |A_i|$.
	
	In the following we describe the score $ \psi_i $ that we propose for the $ k $-Thresholding algorithm. 
	For every test pool $ j \in [m] $ and its outcome $ y_j $, we may look at the \emph{complementary test} pool and its outcome. 
	That is the test pool of the items that were not in the $ j $th test pool. 
	Recall that the number of defective items $ k $ is known (or can be computed, see Section \ref{section:k-is-unkown-or-random}). 
	Therefore, for every test $ j \in [m] $ we can compute the complementary test outcome $\bar{y}_j := k - y_j$ and its test pool which consists of the $ 0 $ entries of the $ j $th test matrix row.
	Accordingly, we define the matrix of complementary tests as $ \bar{A}:= \one_{m \times n} - A $ and the outcome of the complementary tests as $ \bar{y} = k\cdot 1_m - y $.
	Recall that $ x $ is the indicator of defective items and that $ Ax = y $. Note that we have also $ \bar{A}x = \bar{y} $.
	Using this observation, for every item $i \in [n]$, the algorithm calculates its score $\psi_i$ to be the sum of outcomes of tests and complementary tests in which item $i$ is within their test pool. Formally,
	\begin{equation}\label{boosting:eq:estimator_psi}
	\psi_i := \sum_{j=1}^m \left ( A_{ji} y_j + \bar{A}_{ji} \bar{y}_j \right ),
	\end{equation}	
	Note that each score function $ \psi_i $ uses information from $ m $ tests, while $ \phi_i $ uses information from about half that many.
	Additionally and as shortly shown, the distribution of $ \psi_i $ is based on the binomial distribution. 
	As this distribution has been extensively studied, it facilitates the analysis. 
	
	Now we prove Lemma \ref{boosting:lem:top-m-number-of-tests}. 
	The proof of Lemma \ref{boosting:lem:top-k-number-of-tests} is almost identical, so we omit it.
	
	\begin{proof}[Proof of Lemma \ref{boosting:lem:top-m-number-of-tests}]
		We start by describing the distribution of a score function $ \psi_i $ using a binomial random variable.
		Note that each element of the score sum in (\ref{boosting:eq:estimator_psi}) is $ A_{ji} y_j + \bar{A}_{ji} \bar{y}_j $. Namely, that is the number of defective items that are in the same test or complementary-test pool of item $ i $.		
		Therefore, for a non-defective item $ i \in \notD $ and each test $ j \in [m] $ the expression $ A_{ji} y_j + \bar{A}_{ji} \bar{y}_j$ is distributed as the  Binomial random variable $ B\left(k, \nhalf \right) $. 
		Whereas, for a defective item $ i \in \D $ it is distributed as $ 1 + B\left(k-1, \nhalf \right) $.
		Hence, the score function of each non-defective item $i\in [n] \setminus \D$ is distributed as 
		$${\psi_i \sim B(mk, \nhalf)},$$
		whereas each defective item $i \in \D$ score function is distributed as
		$$\psi_i \sim m + B(m(k-1), \nhalf).$$
		
		We need to show that, w.h.p., all $k$ defective items appear among the first $ m $ items sorted by score. 
		We will do so by showing that there exists some value $B\in \mathbb{R}$
		such that, w.h.p, $B$ is smaller than all of the defective-items' scores and is larger than all but potentially $m-k$ non-defective-items' scores.
		
		First, observe that for a non-defective item $i\in [n] \setminus \D$ we have that 
		$$ \mathbb{E}[\psi_i] = \frac{mk}{2} $$ 
		and for defective item $i\in \D$  it is
		$$ \mathbb{E}[\psi_i] = \frac{m(k+1)}{2},$$
		therefore we express $B$ as  
		\begin{equation*}
		\frac{m(k+\beta)}{2}
		\end{equation*}
		for some $\beta \in \left(0,1\right)$. 
		
		For analysis, we will use the following inequality that we prove in the Appendix (see Section \ref{sec:appendix:binomial-tail-bound}).
		\begin{lem}\label{lem:pr-of-bernoulli-not-in-center}
			Consider $ N \in \mathbb{N}$, $ t:=t(N)\in\mathbb{N} $ such that $\sqrt N \le t \le o(N)$, and $X \sim B(N,\nhalf)$. It holds that:
			$$
				\Pr\left [X < \frac{N}{2} - t\right ] = 
				\Pr\left [X >\frac{N}{2} + t\right ] 
				\leq O\left ({\frac{\sqrt N}{t}}  e^{-\frac{2t^2}{N}} \right ).
			$$
		\end{lem}
		Assume that $m > k\lognk$.
		By Lemma \ref{lem:pr-of-bernoulli-not-in-center}, for a non-defective item $ i\in \notD $ we have that 
		$$ 
		\Pr\left [\psi_i \geq \frac{m(k+\beta)}{2}\right ] <O\left(\sqrt\frac{k}{m}e^{-\frac{\beta^2 m}{2k}}\right)
		=o(1)e^{-\frac{\beta^2 m}{2k}},
		$$
		and for defective item $ i \in \D $
		$$ 
		\Pr\left [\psi_i \leq \frac{m(k+\beta)}{2}\right ] < O\left(\sqrt\frac{k}{m}e^{-\frac{(1-\beta)^2 m}{2k}}\right)
		=o(1)e^{-\frac{(1-\beta)^2 m}{2k}}.
		$$
		
		Use Markov's inequality to bound the probability that there are more than $ m - k $ non-defective items with a score larger than $ B $. That is
		\begin{equation} \label{thm1-pr-non-defective-bound}
		o(1) \frac{n-k}{m-k}e^{-\frac{\beta^2 m}{2k}}
		\le o(1) \frac{n}{k}e^{-\frac{\beta^2 m}{2k}}
		\leq o(1)n^{1-\theta}e^{-\frac{\beta^2 m}{2k}}.
		\end{equation}
		Although $ m>k\lognk=O(k \log n) $, note that  in (\ref{thm1-pr-non-defective-bound}) we bound $ m -k $ by $ k $. 
		This costs us a factor of $ 1/\lognk $ in the right hand side expression. 
		However, for our threshold analysis we consider the logarithm of inequality (\ref{thm1-pr-non-defective-bound}), and therefore we lose only a low order term in the exact threshold analysis (see Remark \ref{remark:2k-ssbt-is-also-like-ssbt} for further discussion).
				
		Using a union bound, we get that the probability that there exists a defective item with score below $ B $ is at most:
		\begin{equation} \label{thm1-pr-defective-bound}
			o(1)k e^{-\frac{(1-\beta)^2 m}{2k}}.
		\end{equation}
		Therefore, combining (\ref{thm1-pr-non-defective-bound}) and (\ref{thm1-pr-defective-bound}) we get that if $m$ is such that
		\begin{equation} \label{eq:lem:m-thresholding:objective}
					o(1)n^{1-\theta}e^{-\frac{\beta^2 m}{2k}} + o(1)k e^{-\frac{(1-\beta)^2 m}{2k}} = o(1),
		\end{equation}
		then $ B $ has the desired property.
		
		Note that in order to achieve the above asymptotic bound it suffices that $ m $ will satisfy  
		$ \frac{{\beta^2}m}{2k} > \left(1-\theta\right) \ln n$
		and 
		$ \frac{(1-\beta)^2m}{2k} > \ln k = \theta \ln n$. In other words,
		\begin{equation}\label{thm1-m1-bound}
		m > 2k \cdot \max \left \{ \frac{1-\theta}{\beta^2} , \frac{\theta}{(1-\beta)^2}\right \}\ln n.
		\end{equation} 
		
		We would like to find an $\beta$ which minimizes the above equation's right side. Note that 
		$ \frac{1-\theta}{\beta^2} $ and $\frac{\theta}{(1-\beta)^2}$ are monotonically decreasing and  monotonically increasing in $\beta$, respectively. Therefore, an $ \beta $ for which those two expressions are equal, is optimal.
		We define $\beta(\theta)$ as follows: 
		$$
		\beta(\theta) := 	\left\{
		\begin{array}{ll}
		\nhalf & \mbox{if } \theta=\nhalf \\
		\frac{\theta-1+\sqrt{\theta(1-\theta)}}{2\theta-1} & \mbox{otherwise. }
		\end{array}
		\right.
		$$ 
		As $\frac{1-\theta}{\beta(\theta)^2} = \frac{\theta}{(1-\beta(\theta))^2}$ and $\beta(\theta) \in (0,1)$, the chosen $\beta(\theta)$ is optimal. 
		By assigning $ \alpha = \beta(\theta)^{-1} $ in (\ref{thm1-m1-bound}), we conclude that for  $ m > k \alpha^{2} \ln \frac{n}{k} $
		the set $S$ contains w.h.p. all defective items.
	\end{proof}
	\begin{remark}\label{remark:2k-ssbt-is-also-like-ssbt}
		Note that the analysis in inequality (\ref{thm1-pr-non-defective-bound}) was done for an algorithm that picks only the first $ 2k $ items (instead of first $ m $) sorted by descending order of scores.
		We refer to this algorithm variation as the \emph{$ 2k $-Thresholding}.
		Doing the same analysis but without neglecting the low order term in (\ref{thm1-pr-non-defective-bound}), will yield the same threshold up to a factor of 1-o(1).
		
		In Section \ref{sec:empirical}, we conducted simulations of the algorithms: $ k $-Thresholding, $ m $-Thresholding and $ 2k $-Thresholding (see test set \ref{simulations:testset:bt-ssbt-2kssbt}).		
		For some fixed $ n $ and $ k $, we simulated each algorithm on various values of $ m $.
		We then compare the theoretical thresholds (Lemma \ref{boosting:lem:top-k-number-of-tests} and Lemma \ref{boosting:lem:top-m-number-of-tests}) with the minimal $ m $ value such that these algorithms solved almost all problem instances.		
		The observed success rate of the $ k $-Thresholding and the $ 2k $-Thresholding algorithms agrees with the theoretical threshold given at Lemma \ref{boosting:lem:top-k-number-of-tests} and Lemma \ref{boosting:lem:top-m-number-of-tests}, respectively.
		The $ m $-Thresholding algorithm succeed for much smaller $ m $ values. 
		However, when calculating carefully Lemma \ref{boosting:lem:top-m-number-of-tests}'s threshold and taking in account the low order term in Equation (\ref{thm1-pr-non-defective-bound}), the simulations do agree with the theoretical analysis.
	\end{remark}
	
	\subsection{Proof of Theorem \ref{thm:SSA-with-little-order-term}}
\begin{proof}
	Consider a Subset Select algorithm $ ALG $ as in the theorem statement,
	and a global constant $ C'>0 $ that we set later. 
	Define $ m_1 :=  m - C'\sqrt{m \log n} $ and $ m_2 := m-m_1 $.
	Given the Subset Select algorithm $ ALG $, we propose a new Subset Select algorithm $ ALG' $.
	We will then show that regardless of $ ALG $, the algorithm $ ALG' $ has the relaxed rank property with $ C=0 $, and therefore by Theorem \ref{thm:SSA-rank-condition}, we are done.
	The Subset Select algorithm $ ALG' $,	given a problem instance $ (A,y,k) $,
	runs $ ALG $ on the first $ m_1 $ tests. 
	Namely, it runs $ ALG $ on the problem instance $ (A', y',k) $, where $ A'\in\{0,1\}^{m_1 \times n} $ is a submatrix of $ A $ containing the first $ m_1 $ rows and $ y' $ is the first $ m_1 $ entries of $ y $.
	Then, $ ALG' $ returns the same subset $ \algS \subseteq [n] $ that $ ALG $ returned on the modified problem instance.	
	Note that $ |\algS| \le m_1 $ and that indeed if $ ALG $ solves efficiently the Subset Select problem for $ m_1>M $, then $ ALG' $ solves efficiently the Subset Select problem for $ m=m_1+m_2>(1+o(1))M $.
	
	In the rest of the proof, we use the following lemma that we prove in the Appendix as a corollary of Lemma \ref{SSA:lem:X-in-V} (see Section \ref{sec:apx:lem-space-rank-bound}). 
	Let $u_1, u_2, \ldots, u_l$ be i.i.d. random vectors sampled uniformly from $\{0,1\}^{m_1}$. 
	Denote the vector space spanned by these vectors as $U := \langle u_1, u_2, \ldots, u_l \rangle$ and let $V$ be a deterministic subspace in $ \{0,1\}^{m_1} $.		
	
	\begin{lem} \label{lem:f2-space-rank-bound}
		Let $k_1$ and $k_2$ be integers such that $0 \leq k_1 \leq k_2 \leq {m_1} $. 
		If $\dim\left( V \right) \geq k_1$ then the following holds:
		$$
		\Pr_{u_1,u_2,\ldots,u_l}\left[ 
		\dim\left(
		V \cup U
		\right)
		< k_2
		\right] 
		\leq 
		{l \choose {k_2-k_1-1}}2^{(k_2-1-m_1)(l-(k_2-k_1)+1)}
		$$
	\end{lem}
	We will show that every subset $ S \subseteq [n] $ that was constructed independently of the last $ m_2 $ rows of $ A $ and its cardinality is at most $ m_1 $,  will satisfy $ \rankof{A|_S}=|S| $ \whp.
	This implies that \whp $ \rankof{A|_\algS} = |\algS| $.
	Apart from the fact that $ \algS $ was constructed without observing the last rows, we also use the fact that its cardinality is at most $ m_1 $. 
	For convenience, we assume that $ |\algS|=m_1 $.
	
	Denote by $ B_1 , B_2, \ldots, B_{m_1}, \ldots, B_m $ the rows of the matrix $ A|_{\algS} $.
	Consider the event that the dimension of the space spanned by $ B_1, B_2, \ldots B_{m_1} $ is at least $ m_1-\nicefrac{m_2}{2} $ and denote it by $ E_1 $. Formally, the event $ E_1 $ is
	\begin{equation*}\label{eq:thm-rank-proof:E1}
	\dim \left(\langle B_1,B_2, \ldots, B_{m_1}\right \rangle) \geq m_1-\nicefrac{m_2}{2}.
	\end{equation*}
	Similarly, denote by $ E_2 $ and $ E_3 $ the events:
	\begin{equation*}
	\begin{split}
	\dim \left (\langle B_1,B_2, \ldots, B_{m_1+\nicefrac{m_2}{2}} \rangle\right ) & \geq  m_1-\log m_1,\\
	\dim \left (\langle B_1,B_2, \ldots, B_{m}\right \rangle) & = m_1;
	\end{split}
	\end{equation*}
	respectively. Denote by $ \bar{E}_1, \bar{E}_2 $ and $ \bar{E}_3 $ the complementary events of $ E_1, E_2 $ and $ E_3 $, respectively.
	
	Note that the $ E_3 $ is the event that we are after, i.e., $ \rank\left(A|_{\algS}\right) = m_1 $.
	By the law of total probability we have that:
	\begin{equation*}
	\Pr \left[E_3\right] \geq \Pr [E_1]\cdot \Pr [E_2 | E_1 ] \cdot \Pr [E_3 | E_2].
	\end{equation*}
	Therefore, We will show that the event $ E_3 $ holds w.h.p. by proving that 
	\begin{equation}\label{eq:thm-rank-proof:total-bound}
	\Pr [E_1]\cdot \Pr [E_2 | E_1 ] \cdot \Pr [E_3 | E_2]  = 1 - o(1).
	\end{equation}
	Let $ c > 0 $ be a small enough constant.
	Using lemma \ref{lem:f2-space-rank-bound}, with $ V = \emptyset $ (and therefore $ k_1 = 0 $), $ l = m_1 $ and $ k_2 = m_1-\nicefrac{m_2}{2} $,
	we have that $ \Pr [\bar{E}_1|{\algS}=S_0] $ for any subset $ S_0 $ of cardinality $ m_1 $ is upper bounded by:
	\begin{equation*}
	{m_1 \choose {m_1-\nicefrac{m_2}{2}-1}}2^{(m_1-\nicefrac{m_2}{2}-1-m_1)(m_1-(m_1-\nicefrac{m_2}{2})+1)} \leq
	{m_1 \choose \nicefrac{m_2}{2}+1} 2^{-cm_2^2}
	\end{equation*} 
	Therefore, using a union bound over all possible subsets $S$ of cardinality $m_1$ we have that
	\begin{equation}\label{eq:thm-rank-proof:E1-bound}		  
	\Pr \left[
	E_1
	\right] 
	\geq
	1-
	\sum_{{S_0}}
	\Pr \left[ \bar{E}_1 | \algS = {S_0}\right]
	\geq 
	1-{n \choose m_1}{m_1 \choose \frac{m_2}{2} + 1} 2^{-cm_2^2}
	= 1-o(1).		
	\end{equation}
	The last equality in (\ref{eq:thm-rank-proof:E1-bound}) follows the definition of $ m_1 $ and $ m_2 $. That is, $ m_1 = m -C\sqrt{m \log n} $ and $ m_2 = m-m_1 $.
	
	Note that for every $m_1<j\leq m$ the row $ B_j $ is a uniformly random vector in $ \{0,1\}^{m_1}$ (because these rows are not observed by the algorithm when it constructed  $ \algS $). 
	Next, we give an upper bound to $ \Pr \left[\bar{E}_2|E_1 \right] $.
	Using Lemma \ref{lem:f2-space-rank-bound} with the parameters: $ V = \langle B_1, B_2, \ldots, B_{m_1} \rangle $, $ k_1 = m_1 - \nicefrac{m_2}{2} $, $ l = \nicefrac{m_2}{2} $ and $ k_2 = m_1 - \log m_1 $. 
	We have that $ \Pr \left[\bar{E}_2|E_1 \right] $ is upper bounded by:
	\begin{equation*}
	{\nicefrac{m_2}{2} \choose {\nicefrac{m_2}{2}- \log m_1 - 1}}
		2^{(-\log m_1 - 1)(\log m_1 + 1)} 
	\leq {\nicefrac{m_2}{2} \choose \log m_1 + 1} 
		{m_1}^{-\log {m_1}}
	=o(1)		
	\end{equation*}
	Therefore,
	\begin{equation}\label{eq:thm-rank-proof:E2-bound}
	\Pr [E_2 | E_1] = 1 - \Pr \left[\bar{E}_2|E_1 \right]=1-o(1)
	\end{equation}
	Last, we upper bound $ \Pr \left[ \bar{E}_3 | E_2 \right] $.
	By applying Lemma \ref{lem:f2-space-rank-bound} with the parameters $ V = \langle B_1,B_2, \ldots, B_{m_1+\nicefrac{m_2}{2}} \rangle$, $ l = \nicefrac{m_2}{2} $, $ k_1 = m_1-\log m_1 $ and $ k_2 = m_1 $ we get the upper bound of 
	\begin{equation*}
	{\nicefrac{m_2}{2} \choose {\log m_1 -1 }}
	2^{-(\nicefrac{m_2}{2}-\log m_1+1)}
	\leq {\nicefrac{m_2}{2} \choose \log {m_1} - 1} 
	2^{-c(m_2 - \log m_1)}=o(1).
	\end{equation*}
	Hence,
	\begin{equation}\label{eq:thm-rank-proof:E3-bound}
	\Pr [E_3 | E_2] =
	1-\Pr \left[ \bar{E}_3 | E_2 \right]=1-o(1).
	\end{equation}
	
	Set $ C:=c^{-1} $ and assign (\ref{eq:thm-rank-proof:E1-bound}), (\ref{eq:thm-rank-proof:E2-bound}) and (\ref{eq:thm-rank-proof:E3-bound}) into (\ref{eq:thm-rank-proof:total-bound}). We have that 
	\begin{equation*}
	\Pr[E_3] \ge 1-o(1),
	\end{equation*}
	and the proof follows.
\end{proof}

	\section{Empirical results} \label{sec:empirical}
	
	We performed all simulations on a machine with 2-core Xeon 2.3GHz CPU and 13GB RAM. 
	We implemented all algorithms using Python and Numpy.
	
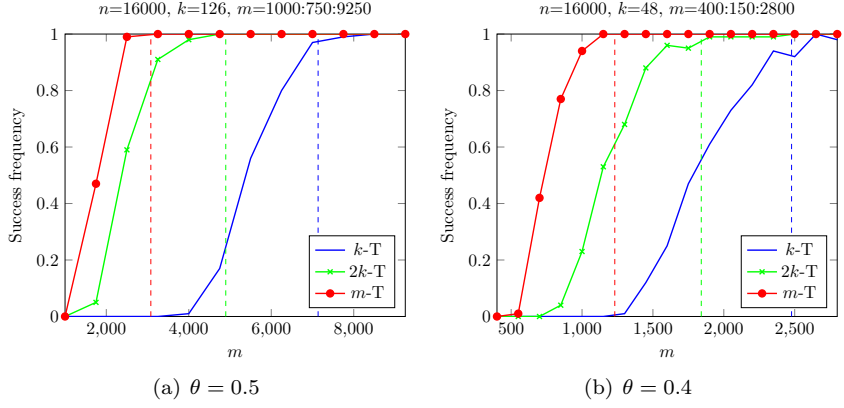
\begin{figure}[!htbp]
	 		\centering
			\subfigure[$ \theta = 0.5 $]
			{
				\label{simulations:testset:bt-ssbt-2kssbt:0.5}
				\begin{tikzpicture}[scale=0.66]
					\begin{axis}[
						title={$n$=$16000$, $k$=$126$, $m$=1000:750:9250},
						xlabel={$m$},
						ylabel={Success frequency},
						ytick={0,0.2,0.4,0.6,0.8,1},
						ymin=0,
						ymax=1,
						xmin=1000,
						xmax=9250,
						legend pos=south east
					]
						\addplot[color=blue, style={thick}] table [x=x, y=y, col sep=comma] {\main/figures/data/bt-k126-n16000.csv};
						\addplot[color=green, mark=x, style={thick}] table [x=x, y=y, col sep=comma] {\main/figures/data/2kssbt-k126-n16000.csv};
						\addplot[color=red, mark=*, style={thick}] table [x=x, y=y, col sep=comma] {\main/figures/data/ssbt-k126-n16000.csv};
						\addplot +[dashed, color=red, mark=none, forget plot] coordinates {(3080, 0) (3080, 1)};
						\addplot +[dashed, color=green, mark=none, forget plot] coordinates {(4897, 0) (4897, 1)};
						\addplot +[dashed, color=blue, mark=none, forget plot] coordinates {(7136, 0) (7136, 1)};
						\legend{$ k $-T,$ 2k $-T, $ m $-T}					
					\end{axis}			
				\end{tikzpicture}
			}
			\subfigure[$ \theta = 0.4 $]
			{
				\label{simulations:testset:bt-ssbt-2kssbt:0.4}
				\begin{tikzpicture}[scale=0.66]
					\begin{axis}[
							title={$n$=$16000$, $k$=$48$, $m$=400:150:2800},
							xlabel={$m$},
							ylabel={Success frequency},
							ytick={0,0.2,0.4,0.6,0.8,1},
							ymin=0,
							ymax=1,
							xmin=400,
							xmax=2800,
							legend pos=south east
						]
						\addplot[color=blue, style={thick}] table [x=x, y=y, col sep=comma] {\main/figures/data/bt-k48-n16000.csv};
						\addplot[color=green, mark=x, style={thick}] table [x=x, y=y, col sep=comma] {\main/figures/data/2kssbt-k48-n16000.csv};
						\addplot[color=red, mark=*, style={thick}] table [x=x, y=y, col sep=comma] {\main/figures/data/ssbt-k48-n16000.csv};
						\addplot +[dashed, color=red, mark=none, forget plot] coordinates {(1231, 0) (1231, 1)};
						\addplot +[dashed, color=green, mark=none, forget plot] coordinates {(1841, 0) (1841, 1)};						
						\addplot +[dashed, color=blue, mark=none, forget plot] coordinates {(2478, 0) (2478, 1)};
						\legend{$ k $-T,$ 2k $-T, $ m $-T}					
					\end{axis}			
				\end{tikzpicture}
			}
			\caption{
				$k$-Thresholding ($ k $-T),  $ 2k $-Thresholding ($ 2k $-T) and $m$-Thresholding ($ m $-T) success frequency out of $ 100 $ trails. 
				The blue, green and red dashed lines indicate the theoretical threshold of these algorithms, in the same order, as discussed in test set \ref{simulations:testset:bt-ssbt-2kssbt}.
			}
		 	\label{figure:bt-ssbt}
	 \end{figure}

	\begin{ts} \label{simulations:testset:bt-ssbt-2kssbt}
		In this set, we simulate the algorithms: 
		$k$-Thresholding (see Algorithm \ref{alg:Basic-Thresholding} in Section \ref{sec:contributions:ssa-algorithms}), 
		$m$-Thresholding (see Section \ref{sec:contributions:ssa-algorithms}) and 
		$ 2k $-Thresholding (see Remark \ref{remark:2k-ssbt-is-also-like-ssbt} in Section \ref{boosting:subsec:basic-thresholding}). 
		We simulated problem instances with $ n=16000 $ items, among them $ k= \lfloor n^{0.5}\rfloor,\lfloor n^{0.4}\rfloor $ are defective, and various $ m $ values.
		For each $ n,k $ and $ m $ combination, we performed $ 100 $ simulations and plotted the success frequency for each algorithm.
		The results for $ k= \lfloor n^{0.5}\rfloor$ are plotted in Figure \ref{simulations:testset:bt-ssbt-2kssbt:0.5}, and for $k=\lfloor n^{0.4}\rfloor $ in 
		Figure \ref{simulations:testset:bt-ssbt-2kssbt:0.4}. 
		
		Additionally, on these plots there are also the theoretical thresholds of these algorithms.
		We use Lemma \ref{boosting:lem:top-k-number-of-tests} for the $ k $-Thresholding, Lemma \ref{boosting:lem:top-m-number-of-tests} for the $ 2k $-Thresholding and the minimal $ m $ value such that
		$$ 
		\min_{\beta\in(0,1)} \max \left\{
			\frac{n-k}{m-k} e^{-\frac{\beta^2m}{2k}},
			k e^{-\frac{(1-\beta)^2m}{2k}}
		\right \} 
		\le1
		$$ 
		for the $ m $-Thresholding (see discussion in Remark \ref{remark:2k-ssbt-is-also-like-ssbt}). 
		These figures show that the simulation results agree with our theoretical calculations already for a small instance as $ n=16000 $.
	\end{ts}
	 
\begin{figure}[!htbp]
	\centering
	    \begin{tikzpicture}[scale=0.66]
	\pgfplotsset{grid style={dashed,gray}}
			\begin{axis}[
	       name=ax1,	
	title={$n$=$1000$, $k$=$100$, $m$=250:50:950},
	xlabel={$m$},
	ylabel={Success frequency},
	ytick={0,0.2,0.4,0.6,0.8,1},
	ymin=0,
	ymax=1,
	xmin=250,
	xmax=950,
        legend style={font=\small, at={(0.3,0-0.2)},anchor=north west,legend columns=6},
	]

	\addplot[color=red, mark=*, dashed] table [x=x, y=y, col sep=comma] {\main/figures/data/mp-k100-n1000.csv};
	\addplot[color=red, mark=*, style={thick}] table [x=x, y=y, col sep=comma] {\main/figures/data/ssmp-k100-n1000.csv};
		
	\addplot[color=blue, mark=x, dashed] table [x=x, y=y, col sep=comma] {\main/figures/data/ompq-k100-n1000.csv};
	\addplot[color=blue, mark=x, style={thick}] table [x=x, y=y, col sep=comma] {\main/figures/data/ssompq-k100-n1000.csv};

	\addplot[color=green, mark=square, dashed] table [x=x, y=y, col sep=comma] {\main/figures/data/k_thresholding-n=1000-k=100-m=250_50_1000-trais=100.csv};
	\addplot[color=green, mark=square, style={thick}] table [x=x, y=y, col sep=comma] {\main/figures/data/k_t_tt-n=1000-k=100-m=50_50_1050-trais=100.csv};

				\legend{IT,IT-TT, Q-OMP, Q-OMP-TT, $ k $-T, $ k $-T-TT}					
	
	\end{axis}

	\begin{axis}[
	at={(ax1.south east)},
	        legend style={font=\small, at={(0,-0.1)},anchor=mid west,legend columns=3},
	        xshift=2cm,
title={$n$=$1000$, $k$=$160$, $m$=250:50:950},
xlabel={$m$},
ylabel={Success frequency},
ytick={0,0.2,0.4,0.6,0.8,1},
ymin=0,
ymax=1,
xmin=250,
xmax=950,
legend pos=south east
]
\addplot[color=red, mark=*, dashed] table [x=x, y=y, col sep=comma] {\main/figures/data/mp-k160-n1000.csv};
\addplot[color=red, mark=*, style={thick}] table [x=x, y=y, col sep=comma] {\main/figures/data/ssmp-k160-n1000.csv};

\addplot[color=blue, mark=x, dashed] table [x=x, y=y, col sep=comma] {\main/figures/data/ompq-k160-n1000.csv};
\addplot[color=blue, mark=x, style={thick}] table [x=x, y=y, col sep=comma] {\main/figures/data/ssompq-k160-n1000.csv};

\addplot[color=green, mark=square, dashed] table [x=x, y=y, col sep=comma] {\main/figures/data/k_thresholding-n=1000-k=160-m=250_50_1000-trais=100.csv};
\addplot[color=green, mark=square, style={thick}] table [x=x, y=y, col sep=comma] {\main/figures/data/k_t_tt-n=1000-k=160-m=50_50_1050-trais=100.csv};

\end{axis}		
	\end{tikzpicture}  
	\caption{Iterative-Thresholding (IT), Iterative-Thresholding-Then-Thresholding (IT-TT), Q-OMP, Q-OMP-Then-Thresholding (Q-OMP-TT), $ k $-Thresholding ($ k $-T) and $ k $-Thresholding-Then-Thresholding ($ k $-T-TT) success frequency out of $ 100 $ trails. See test set \ref{simulations:testset:tbt-framework} for details.}
	\label{figure:ir-ittbt}
\end{figure}
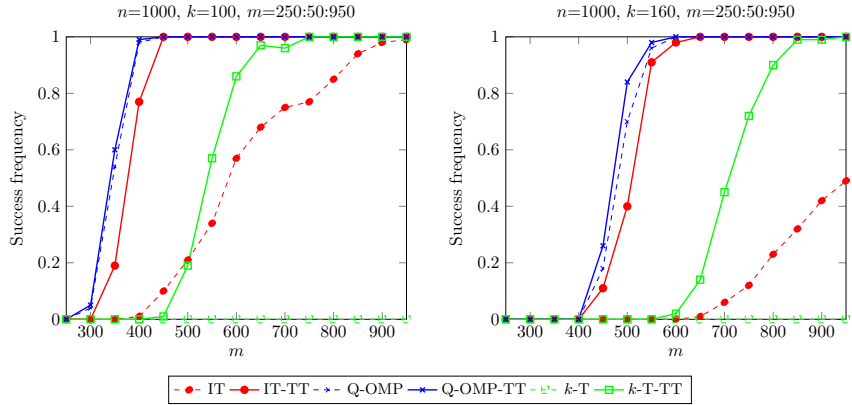

	\begin{ts}\label{simulations:testset:tbt-framework}
		In this test set, we simulate QGT algorithms and their adaption to the Subset Select problem using our proposed framework (defined in Section \ref{sec:contributions:ssa-algorithms}).
		We have conducted simulations and documented the success frequency of $k$-Thresholding, Iterative-Thresholding (see Algorithm \ref{alg:Iterative-Thresholding}) and {Q-OMP} (see Algorithm \ref{alg:Q-OMP}).
		We compare each of these algorithms with its resulting Subset Select algorithm.
		We tried to simulate this test set on problem instances with $ n=16000 $ (as in test-set \ref{simulations:testset:bt-ssbt-2kssbt}), however Q-OMP algorithm didn't halt in a reasonable time. 
		Therefore, we used smaller problem instances.
		We simulated problem instances with $ n=1000 $ items, among them $ k= 100,160$ are defective, and various $ m $ values.
		We set the $ E $ parameter in the Q-OMP algorithm to $ E = k+30 $.
		For each $ n,k $ and $ m $ combination, we performed $ 100 $ simulations and plotted the success frequency for each algorithm (see Figure \ref{figure:ir-ittbt}).
		A success for a QGT algorithm is to recover exactly $ \D$ and for the Subset Select algorithm success is when its outcome contains $ \D $.

		Note that the difference between ($ k $-Thresholding)-Then-Thresholding and $ m $-Thresholding is that the first treats the $ k $-Thresholding algorithm as a black-box and extends it without changing the algorithm behavior (see test set \ref{simulations:testset:compare-qgt-algs} for a comparison between the two).
		
		Observe that while the $ k $-Thresholding and the Iterative-Threshold algorithms improve significantly by the framework, the Q-OMP algorithm does not.
	\end{ts}

	\begin{algorithm}[!htbp]
		\caption{Iterative-Thresholding (QGT algorithm)}\label{alg:Iterative-Thresholding}
		\begin{algorithmic}
			\STATE \textbf{Input:} $ A,y,k $
			\STATE \textbf{Init:} $ \algSz \leftarrow \emptyset $, $ y_0 \leftarrow y $
			\begin{enumerate}
				\item For $ t = 1 $ until $ t=k $:
				\begin{enumerate}
					\item For every item $i \in [n] \setminus S $, calculate its score $\phi_{t,i} := \langle A_i, y_{t-1} \rangle / \|A_i\|_1.$
					\item Set $ s_t \leftarrow {\arg\max}_{i\in [n]\setminus \algStmo} \phi_{t,i} $.
					\item Set $ \algSt \leftarrow \algStmo \cup \{s_t\} $.
					\item Set $ y_t \leftarrow y_{t-1} - A_{s_t} $.					
				\end{enumerate}
			\end{enumerate}
			\STATE \textbf{Output:} $ \algSk $
		\end{algorithmic}
	\end{algorithm}
	\begin{algorithm}[!htbp]
		\caption{\cite{sparrer2013discrete,sparrer2015soft} { } Q-OMP (QGT algorithm)}\label{alg:Q-OMP}
			\begin{algorithmic}
			\STATE \textbf{Input:} $ A,y,k $
			\STATE \textbf{Init:} $ \algSz \leftarrow \emptyset $, $ y_0 \leftarrow y $
			\begin{enumerate}
				\item For $ t = 1 $ until $ t=E $:
				\begin{enumerate}
					\item For every item $i \in [n] \setminus S $, calculate its score $\phi_{t,i} := \langle A_i, y_{t-1} \rangle / \|A_i\|_1.$
					\item Set $ s_t \leftarrow {\arg\max}_{i\in [n]\setminus \algStmo} \phi_{t,i} $.
					\item Set $ \algSt \leftarrow \algStmo \cup \{s_t\} $.
					\item Set $ z_t \leftarrow \text{argmin}_{z'\in \mathbb{R}^t_+} \| (A|_\algSt)z' - y \|_2 $.
					\item Define $ x_t \in \{0,1\}^n$ such that 
					$$(x_t)_i:= \left\{
					\begin{array}{ll}
					1 & \mbox{if } i=s_j \in \algSt \mbox{ and } z'_j>0.6 \\
					0 & \mbox{otherwise. }
					\end{array}
					\right. $$
					\item Set $ y_t \leftarrow y_{0} - Ax_t $. If $ \|y_t\| = 0 $, halt.
				\end{enumerate}
			\end{enumerate}				
			\STATE \textbf{Output:} $ \left \{i\in [n] \prmid (x_t)_i = 1\right \}$
		\end{algorithmic}
	\end{algorithm}

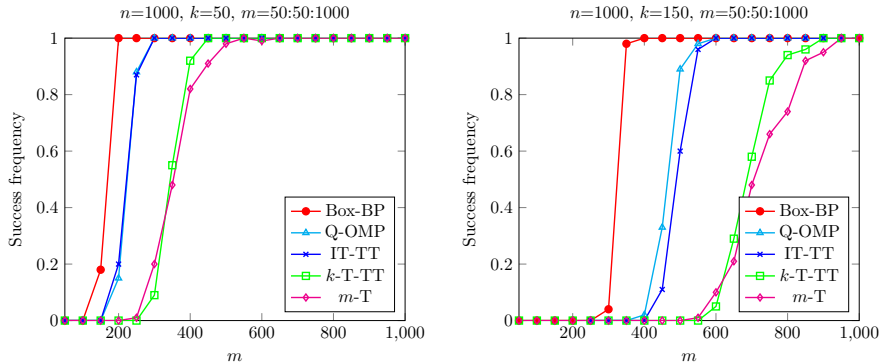
\begin{figure}[!htbp]
		\centering
		\subfigure{
			\label{simulations:testset:mp-ssmp:1000-100}
			\begin{tikzpicture}[scale=0.66]
			\begin{axis}[
				title={$n$=$1000$, $k$=$50$, $m$=50:50:1000},
				xlabel={$m$},
				ylabel={Success frequency},
				ytick={0,0.2,0.4,0.6,0.8,1},
				ymin=0,
				ymax=1,
				xmin=50,
				xmax=1000,
				legend pos=south east
			]
							\addplot[color=red, mark=*, style={thick}] table [x=x, y=y, col sep=comma] {\main/figures/data/boxed_bp-n=1000-k=50-m=50_50_1050-trais=100.csv};
				\addplot[color=cyan, mark=triangle, style={thick}] table [x=x, y=y, col sep=comma] {\main/figures/data/Q_OMP-n=1000-k=50-m=50_50_1050-trais=100.csv};

				\addplot[color=blue, mark=x, style={thick}] table [x=x, y=y, col sep=comma] {\main/figures/data/iterative_threshold_then_threshold-n=1000-k=50-m=50_50_1050-trais=100.csv};			
				
				\addplot[color=green, mark=square, style={thick}] table [x=x, y=y, col sep=comma] {\main/figures/data/k_t_tt-n=1000-k=50-m=50_50_1050-trais=100.csv};
				
				\addplot[color=magenta, mark=diamond, style={thick}] table [x=x, y=y, col sep=comma] {\main/figures/data/m_thresholding-n=1000-k=50-m=50_50_1050-trais=100.csv};
				
				\legend{Box-BP,Q-OMP, IT-TT, $ k $-T-TT, $ m $-T}					
			
			\end{axis}			
			\end{tikzpicture}
		}
		\subfigure
		{
			\label{simulations:testset:mp-ssmp:1000-160}
			\begin{tikzpicture}[scale=0.66]
			\begin{axis}[
			title={$n$=$1000$, $k$=$150$, $m$=50:50:1000},
			xlabel={$m$},
			ylabel={Success frequency},
			ytick={0,0.2,0.4,0.6,0.8,1},
			ymin=0,
			ymax=1,
			xmin=50,
			xmax=1000,
			legend pos=south east
			]
								\addplot[color=red, mark=*, style={thick}] table [x=x, y=y, col sep=comma] {\main/figures/data/boxed_bp-n=1000-k=150-m=50_50_1050-trais=100.csv};
				\addplot[color=cyan, mark=triangle, style={thick}] table [x=x, y=y, col sep=comma] {\main/figures/data/Q_OMP-n=1000-k=150-m=50_50_1050-trais=100.csv};

				\addplot[color=blue, mark=x, style={thick}] table [x=x, y=y, col sep=comma] {\main/figures/data/iterative_threshold_then_threshold-n=1000-k=150-m=50_50_1050-trais=100.csv};			
				
				\addplot[color=green, mark=square, style={thick}] table [x=x, y=y, col sep=comma] {\main/figures/data/k_t_tt-n=1000-k=150-m=50_50_1050-trais=100.csv};
				
				\addplot[color=magenta, mark=diamond, style={thick}] table [x=x, y=y, col sep=comma] {\main/figures/data/m_thresholding-n=1000-k=150-m=50_50_1050-trais=100.csv};
				
				\legend{Box-BP,Q-OMP, IT-TT, $ k $-T-TT, $ m $-T}					
			
			\end{axis}		
			\end{tikzpicture}
		}
		\caption{Box-BP, Q-OMP, Iterative-Thresholding-Then-Thresholding (IT-TT), $ m $-Thresholding and $ k $-Thresholding-Then-Thresholding ($ k $-T-TT) success frequency out of $ 100 $ trails. See test set \ref{simulations:testset:compare-qgt-algs} for details.}
		\label{figure:general-algs}
	\end{figure}

	\begin{ts}\label{simulations:testset:compare-qgt-algs}
		In this test set we compare between the performance of different algorithms.
		We have conducted simulations and documented the success frequency of $ k $-Thresholding-Then-Thresholding, $m$-Thresholding, Iterative-Thresholding-Then-Thresholding, Q-OMP and Box-BP (see Section\ref{sec:related-works}).
		We tried to simulate this test set on problem instances with $ n=16000 $ (as in test-set \ref{simulations:testset:bt-ssbt-2kssbt}), however Bin-BP crushed due to a lack of RAM. 
		Therefore, we used smaller problem instances.
		We simulated problem instances with $ n=1000 $ items, among them $ k= 50,150$ are defective, and various $ m $ values.
		We set the $ E $ parameter in the Q-OMP algorithm to $ E = k+30 $.
		For each $ n,k $ and $ m $ combination, we performed $ 100 $ simulations and plotted the success frequency for each algorithm (see Figure \ref{figure:general-algs}).
		
		Observe that the $ m $-Thresholding and the $ k $-Thresholding-Then-Thresholding algorithms performed similarly.
	\end{ts}
	 
	\FloatBarrier

\section*{Acknowledgement}
This research was supported in part by The Israel Science Foundation (grant No. 1388/16).
We would like to thank Tal Amir, Yonina Eldar, Boaz Nadler, Danny Vilenchik, and Ofer Zeitouni for helpful discussions and pointers to references.

	\bibliography{bibfile}
	
	\appendix

	\section{Appendix}
			\subsection{Preliminaries}\label{sec:apx:pre}
				In this section we give proofs to known inequalities that we use.
				Let $ N, M $ be integers such that $ M < N $.
				Using the Sterling inequality, stated as  
				$$ 
				\sqrt{2\pi}N^{N+\frac{1}{2}}e^{-N}\leq N! \leq eN^{N+\frac{1}{2}}e^{-N}, 
				$$ 
				we get the following:
				\begin{lem}
					\label{lem:appendix:binomial}
					For $ N $ and $ M $ as the above, it holds that
					$$
					{N \choose M} 
					\le  \left(\frac{eN}{M} \right)^M
					$$
				\end{lem}
				\begin{proof}
					
					$$
					{N \choose M} 
					\leq \frac{N^M}{M!} 
					\leq \frac{N^M}{\sqrt{2\pi M}(M/e)^M} 
					\leq \left (\frac{eN}{M}\right )^M
					$$ 
				\end{proof}
				Consider the Binomial random variable $ X \sim B\left(N, \nicefrac{1}{2}\right) $.
				Using the Sterling inequality we get also that
				\begin{equation}\label{eq:apendix:choose}
				\begin{split}
				\Pr [X=M] 
				& = {N \choose M} 2^{-N} \\
				& = \frac{N!}{M!(N-M)!} 2^{-N} \\
				& \leq \frac{e}{2\pi} 
				\sqrt{\frac{N}{M(N-M)}}
				\left(\frac{N}{M}\right)^M
				\left(\frac{N}{N-M}\right)^{N-M}
				2^{-N}\\
				& = \frac{e}{2\pi} 
				\sqrt{\frac{N}{M(N-M)}}
				\left(\frac{N}{2M}\right)^M
				\left(\frac{N}{2(N-M)}\right)^{N-M}.
				\end{split}
				\end{equation}
				
				\begin{lem}\label{lem:binomial-2n-n}
					Let $ X_1,X_2, \sim B\left (N, \nicefrac{1}{2}\right ) $. The following holds:
					$$
					\Pr\left[ X_1 = X_2 \right]= {2N \choose N}2^{-2N} \leq \frac{e}{ \pi\sqrt{2 N}}
					$$
				\end{lem}
				\begin{proof}
					Define the random variable $ \bar{X}_1 = N - X_1 $. Note that $ \bar{X}_1 $ is also distributed as $ B\left( N, \nicefrac{1}{2} \right) $ and that $ X_1 = X_2 $ if and only if $ \bar{X}_1 + X_2 = N $.
					Therefore, for $ \bar{X}_1 + X_2 =  Y \sim B\left (2N, \nicefrac{1}{2}\right ) $ we have
					$$
					\Pr\left[ Y = N \right] = \Pr [X_1 = X_2].
					$$
					By (\ref{eq:apendix:choose}) we have that:
					$$ 
					\Pr [Y=N]
					= {2N \choose N}2^{-2N}
					\leq \frac{e}{2\pi} 
					\sqrt{\frac{2N}{N^2}}
					\left(\frac{2N}{2N}\right)^N
					\left(\frac{2N}{2N}\right)^N
					=  \frac{e}{ \pi\sqrt{2 N}}
					$$
				\end{proof}
			\subsection{Information theoretic lower bound} \label{appendix:tlb}
				We now provide a proof sketch for the lower bound in equation \ref{intro:eq:info-theortic-bound-sublinear}.
				We need at least $ \log {n \choose k} \sim k\lognk$ bits to express all possible subsets $ \D \subseteq [n] $ of cardinality $ k $.
				A test outcome can have $ k+1 $ different values. 
				However, a typical test outcome is of distance $ O (\sqrt k ) $ from its expectation.
				Therefore to encode all of the typical outcomes we need $\sim \frac{1}{2}\log {k} $ bits.
				(Sometimes a test solution is not in the typical range, but this has negligible effect on the bounds. Further details omitted.)
				We conclude that if the number of tests is below $ \sim 2k \frac{\lognk}{\log k} $ there is not enough information to express all possible subsets $ \D $. 
			\subsection{Information theoretic threshold} \label{appendix:tub}
				\begin{lem}
					\label{intro:lem:tup}
						For QGT problem in the sublinear regime, the information-theoretic threshold on the number of tests is
						\begin{equation}
						\Mqgt < \tub.
						\end{equation}    
				\end{lem}     					
				\begin{proof}[Proof of Lemma \ref{intro:lem:tup}]
					\newcommand{\far}{far}		
					Recall that $ k = n^\theta $ for some constant $ \theta \in (0,1) $.					
					In the following, because $ A $ is a random Bernoulli matrix, we assume w.l.o.g that $ x $ is a fixed vector of Hamming weight $ k $.					
					Let $ l \in [k-1]$ be an integer and let $ z \in \{0,1\}^n $ be a vector of Hamming weight $ k $. We say that $ z $ is \emph{$ l $-\far} from $ x $ if $ x $ and $ z $ are of Hamming distance $ 2l $ (i.e., $ z $ has $ l $ different coordinates $ i\in [n] $ such that $ z_i = 1 $ but $ x_i = 0 $ ). Denote by $ \mathcal{L}_l $ the set of all vectors $ v \in \{0,1\}^n $ of Hamming weight $ k $ that are {$ l $-\far} from $ x $. By using a union bound, the probability that there exists a solution other than $ x $ is upper bounded by:
					\begin{equation}\label{eq:lem:tup:mission}
					\sum_{l=1}^{k} \sum_{z \in \mathcal{L}_l} \Pr [Az = y]
					\end{equation}
					
					Next, we provide an upper bound on the probability that a vector $ z \in \mathcal{L}_l $ is a solution.
					Note that the event $ Az = y $ occurs if for every test $ j\in[m] $, it holds that $ (Ax)_j = (Az)_j $. 
					Because $ A $ is a random Bernoulli matrix, the probability of this event to occur is exactly as that $ m $ times, two Binomial random variables $ X_1,X_2 \sim B(l, \nhalf) $ have the same value.
					By Lemma \ref{lem:binomial-2n-n}, we have that each of these $ m $ events happens with probability of at most $ \frac{e}{\pi \sqrt{2l}} $. Hence, for a fixed $ z\in \mathcal{L}_l $ we have
					$$ 
					\Pr_A \left[ Az=y \right] \leq \left(\frac{e}{\pi \sqrt{2l}} \right)^m.
					$$
					From the above and because $ \mathcal{L}_l $ is of cardinality $ {k \choose k-l}{{n - k} \choose {l}} $, we have that for each $ l \in [k-1] $ it holds that 
					\begin{equation}\label{eq:lem-tub:l-overlap-bound}
					\sum_{
						z \in \mathcal{L}_l
					}
					\Pr \left[Az=y \right] 
					\leq 
					{k \choose k-l}
					{{n - k} \choose {l}} 
					\left(\frac{e}{\pi \sqrt{2l}} \right)^m
					\leq 
					{k \choose l}
					{{n} \choose {l}} 
					\left(\frac{e}{\pi \sqrt{2l}} \right)^m.
					\end{equation}
					From Lemma \ref{lem:appendix:binomial} we have that 
					\begin{equation*}
						{k \choose l}{{n} \choose {l}} 
						\le \left(\frac{ek}{l}\right)^l \left(\frac{en}{l}\right)^l
						=\left(\frac{e^2kn}{l^2}\right)^l.
					\end{equation*}		
					We will show that for every $ l\in [k] $ it holds that
					\begin{equation}\label{appendix:eq:lem-tub:sub-mission}
					k \sum_{
						z \in \mathcal{L}_l
					}
					\Pr \left[Az=y \right] 
					\leq 
						k\left(\frac{e^2kn}{l^2}\right)^l \left(\frac{e}{\pi \sqrt{2l}} \right)^m = o(1)
					\end{equation}
					Denote by $ l_0 \in [k] $ the index which maximizes (\ref{appendix:eq:lem-tub:sub-mission}). 
					Note that this expression for $ l_0 $ bounds from above (\ref{eq:lem:tup:mission}) and therefore by proving this we are done.
					
					We first show that there exists $ \alpha \in (0,1) $ that depends only on $ \theta $ such that for all $ 1\le l \le \alpha \frac{k}{\log k} $ equation (\ref{appendix:eq:lem-tub:sub-mission}) holds.
					
					Assuming that $ m > 2k\frac{\log \frac{e^2 n}{k}}{\log k} = (1+o(1))2k\frac{\lognk}{\log k}$,
					\begin{equation*}
						\begin{split}
						k\cdot \left(\frac{e^2kn}{l^2}\right)^l \cdot \left(\frac{e}{\pi \sqrt{2l}} \right)^m 
						& \le k \cdot \left(e^2 n^{1+\theta} \right)^l \cdot 2^{-\frac{m}{2}} \\
						& \le k \cdot \left(e^2 n^{1+\theta} \right)^l \cdot \left(e^2 n^{1-\theta}\right)^{-\frac{k}{\log k}} \\
						& \le k e^{2(l-\frac{k}{\log k})} \cdot \left(n^{1+\theta} \right)^{l-\frac{1-\theta}{1+\theta}\frac{k}{\log k}}.
						\end{split}
					\end{equation*}
					Note that there exists $ \alpha \in (0,1) $ that depends only on $ \theta $ such that for all $ 1\le l \le \alpha \frac{k}{\log k} $ we have $ l-\frac{k}{\log k} < 2\log k $ and such $ l-\frac{1-\theta}{1+\theta}\frac{k}{\log k} < 0 $. We are left to show that (\ref{appendix:eq:lem-tub:sub-mission}) holds for $ \alpha \frac{k}{\log k} \le l \le k $.
					
					Assume that
					$$ m > 2k\frac{\log \left(\frac{{\log (k)} ^2 e^2}{\alpha^2} \frac{n}{k}\right)}{\log \left(\frac{\alpha k}{\log k}\right)} = (1+o(1))2k\frac{\lognk}{\log k}.$$
					We have that
					\begin{equation*}
					\begin{split}
						k \cdot \left(\frac{e^2kn}{l^2}\right)^l \cdot \left(\frac{e}{\pi \sqrt{2l}} \right)^m 
						& \le k \cdot \left(\frac{{\log (k)} ^2}{\alpha^2} \frac{e^2kn}{k^2}\right)^l \cdot \left(\frac{\log k}{\alpha k}\frac{e^2}{\pi^2 2 } \right)^\frac{m}{2}  \\
						& \le k \cdot 2^{-\frac{m}{2}} \left(\frac{{\log (k)} ^2}{\alpha^2} \frac{e^2kn}{k^2}\right)^{l-k} \\
						& = o(1).
					\end{split}
					\end{equation*}	
				\end{proof}
	\subsection{Binomial random variable tail bound}
		\label{sec:appendix:binomial-tail-bound}
		\begin{lem} \label{lem:binomial-n/2+k}
			Let $ N, t $ be integers such that $ t < \frac{N}{2} $ and consider the Binomial random variable $ X \sim B\left(N, \nicefrac{1}{2}\right) $.
			We have that:
			$$
			\Pr\left[ X = \frac{N}{2} + t \right] 
			\leq 
			\frac{e}{2\pi} 
			\sqrt{\frac{N}{\frac{N^2}{4}-t^2}} e^{-\frac{2t^2}{N}}.
			$$
		\end{lem}
		\begin{proof}
			Recall that $ h(p):= -p \log(p)-(1-p)\log(1-p) $ is the binary entropy function.
			By (\ref{eq:apendix:choose}) we have that:
			\begin{equation*}
			\begin{split}
			\Pr [X=\frac{N}{2} + t] 
			& \leq \frac{e}{2\pi} 
			\sqrt{\frac{N}{\left(\frac{N}{2} + t\right)\left(\frac{N}{2} - t\right)}} \\			
			& \cdot \left(\frac{N}{\frac{N}{2} + t}\right)^{\frac{N}{2} + t}
			\left(\frac{N}{\frac{N}{2} - t}\right)^{\frac{N}{2} - t}2^{-N} \\
			& = \frac{e}{2\pi} 
			\sqrt{\frac{N}{\frac{N^2}{4} - t^2}}
			2^{N(h(\frac{1}{2}+\frac{t}{N}) -1)}.
			\end{split}
			\end{equation*}
			The Taylor series of the binary entropy function in a neighborhood of $ \frac{1}{2} $ is 
			\begin{equation*}
			h(p)=1 - \frac{1}{2\ln 2} \sum^{\infty}_{i=1} \frac{(1-2p)^{2i}}{i(2i-1)}.
			\end{equation*}
			Hence,
			\begin{equation*}
			h\left (\frac{1}{2}+\frac{t}{N}\right )-1
			= - \frac{1}{2\ln 2} \sum^{\infty}_{i=1} \left (\frac{2t}{N}\right )^{2i}\frac{1}{i(2i-1)}
			\le -\frac{1}{\ln 2}\cdot \frac{2t^2}{N^2}.	
			\end{equation*}			
			Therefore, the proof follows.
		\end{proof}
	
		The proof of Lemma \ref{lem:pr-of-bernoulli-not-in-center} follows by applying Lemma \ref{lem:binomial-n/2+k} to the following crude inequality (for a tighter bound see \cite{klar2000bounds})		
		\begin{equation*}
			\Pr[X\ge \frac{N}{2}+t] \le O\left(\frac{N}{t}\right)\Pr[X= \frac{N}{2}+t].
		\end{equation*}
	\subsection{Proof of Lemma \ref{lem:f2-space-rank-bound}} \label{sec:apx:lem-space-rank-bound}
		In the proof of Lemma \ref{lem:f2-space-rank-bound} we denote $ M := m_1 $ for convenience.
		\begin{proof}[Proof of Lemma \ref{lem:f2-space-rank-bound}]
			Recall that by Lemma \ref{SSA:lem:X-in-V} the probability of a uniformly random binary vector $ X \sim \{0,1\}^M $ to belong a deterministic subspace $ V \subseteq \mathbb{R}^M $ with dimension of at most $ t $ is bounded above by
			\begin{equation}\label{lem_proof_f2_vec}		
			\Pr[x \in V] \le 2^{t-M}.
			\end{equation}
			Next, observe that 
			the event 
			$ \dim\left(V \cup U\right) < k_2 $
			occurs if and only if there exists a subset $ S \subseteq \{ u_1 ,u_2, \ldots, u_l\} $ of cardinality $ l - (k_2 - k1) + 1 $ of vectors that are linearly dependent on other vectors. We may assume w.l.o.g. that the set $ S $ is such that for each $ u_i \in S $ it holds that $ u_i \in V \cup \langle u_1,u_2, \ldots, u_{i-1} \rangle $. Therefore we have that:
			\begin{equation}\label{lem_proof_f2_sums}
			\Pr\left[ 
			\dim\left(
			V \cup U
			\right)
			< k_2
			\right] 
			\leq
			\sum_{S}\Pr[\forall u_i \in S \text{ } u_i \in V \cup \langle u_1, u_2, \ldots, u_{i-1}\rangle],
			\end{equation}
			where the sum is taken over all subsets $ S $ of the required cardinality.
			We only need to bound events such that $ \dim(V \cup \langle u_1, u_2, \ldots, u_{i-1}\rangle) \le k_2 -1 $.
			Therefore from \ref{lem_proof_f2_vec}, 
			$$
			\Pr[\forall u_i \in S \text{ } u_i \in V \cup \langle u_1, u_2, \ldots, u_{i-1}\rangle] \le (2^{k_2-1-M})^{|S|}=2^{(k_2-1-M)(l-(k_2-k_1)+1)}
			$$
			We sum over exactly $ l \choose {l-(k_2-k_1)+1} $ possible subsets, and the proof follows.
		\end{proof}

	\subsection{The number of defective items is unknown} \label{section:k-is-unkown-or-random}
			
	In this work, we assume that $ k $ is given as a problem parameter. 
	However, we can omit this assumption because the algorithm can calculate $ k $ \whp from the tests outcome.
	Recall that every item (and specifically a defective item) is in a test pool with probability $ \nhalf $. 
	Therefore, every test outcome $ y_j $ is essentially an i.i.d.\@ Binomial random variable $ B\left(k, \nhalf\right) $. 
	Consider the average of these tests outcome $ Y = \|y\|_1 / m $.
	The expected value of $ Y $ is $ k/2 $.
	Note that the variance of $ Y $ is $ \frac{k}{4m} $.
	By Chebyshev's  inequality, for any non-negative integer $ t \in \mathbb{Z}_+ $ we have that 
	\begin{equation} \label{eq-last}
		\Pr\left [\left |Y-\frac{k}{2}\right | > t+\frac{1}{2} \right] \le \frac{k}{4m}\left (\frac{1}{2}+t\right )^{-2}.
	\end{equation}
	If $ m $ is much larger than $ k $, then the above inequality can be used with $ t=0 $. 
	In this case, $ k=\text{round}(2Y) $ \whp.
	
	The $ k $ parameter can be computed also when $ m<k $.
	Recall that the information theoretical threshold $ \Mqgt $ is such that $ m>O\left( \frac{k}{\log k} \right) $.
	Therefore, $ \frac{k}{4m} < O(\log k) $.
	For $ t = O(\sqrt{\log k}) $ we have that Equation (\ref{eq-last}) holds \whp.
	Suppose $ ALG $ is an algorithm that expects $ k $ as a parameter.
	Calculate $ 2Y $, and run $ ALG $ with each possible integer $ k = k' $ such that $ 2Y -t <k'<2Y +t $.
	If for some $ k' $,  $ ALG $ returns a subset $ S $ of cardinality $ k' $ such that $ A\one_S = y $ then return $ S $.
	It is possible to extend Lemma \ref{intro:lem:tup} (Section \ref{appendix:tub}) to show that if the algorithm returns such an outcome, then  \whp it is indeed the defective items $ \D $. Further details omitted.

\end{document}